\documentclass[12pt]{article}

\usepackage{amsfonts}
\usepackage{amsmath}
\usepackage{amssymb}
\usepackage[dvipdfmx]{graphicx}
\usepackage{subcaption}
\usepackage{color}

\newcommand{\dip}{{d_H}}
\newcommand{\bomega}{{\vec{\omega}}}
\begin{document}

\begin{titlepage}
\begin{flushleft}
{\bf OUJ-FTC-7}\\
{\bf OCHA-PP-369}\\
\end{flushleft}
	
\vskip 5mm
	
\begin{center}
		{\Large Analytical Study of Anomalous Diffusion}\\
		{\Large by Randomly Modulated Dipole}
		\vskip 1cm
		{\large S. Katagiri$^1$, Y. Matsuo$^2$, Y. Matsuoka$^{1\dagger}$ and A. Sugamoto$^3$}
		\vskip 1cm
		{\it $^1$ Nature and Environment, Faculty of Liberal Arts, The Open University of Japan, Chiba 261-8586, Japan}\\
 {\it $^2$ Department of Physics, 
 	Trans-scale Quantum Science Institute,  Mathematics and Informatics Center,
 	University of Tokyo, Hongo 7-3-1, Bunkyo-ku, Tokyo 113-0033, Japan}\\
 		{\it $^3$ Department of Physics, Graduate School of Humanities and Sciences, Ochanomizu University, 2-1-1 Otsuka, Bunkyo-ku, Tokyo 112-8610, Japan}
	\end{center}
	\vfill

\begin{flushleft} 
$^\dagger$Email: machia1805@gmail.com  
\end{flushleft}

\begin{abstract}
This paper derives the Fokker-Planck (FP) equation for a particle moving in potential by a randomly modulated dipole. The FP equation describes the anomalous diffusion observed in the companion paper \cite{AFKMMSSYY} and breaks the conservation of the total probability at the singularity by the dipole. It also shows anisotropic diffusion, which is typical in fluid turbulence. We need to modify the probability density by introducing a mechanism to recover the particle. After the modification, the latent fractal dimension matches with the results of the companion paper. We hope that our model gives a new example of fractional diffusion, where the random singularity triggers fractionality.
\end{abstract}
	\vfill
\end{titlepage}

\setcounter{footnote}{0}

\section{Introduction}
In this paper, we analyze a random process of a particle in $D$-dimensions described by,
\begin{equation}\label{eom}
	\frac{d\vec r}{dt}= \frac{\dip}{r^D}\left(\vec a-D\frac{ (\vec a\cdot\vec r)\, \vec r}{r^2}\right),\quad \vec r\in \mathbb{R}^D.
\end{equation}
or
\begin{equation}
	\frac{dr_i}{dt}=\dip \sum_j V^{(D)}_{ij} a_j, \quad V^{(D)}_{ij}=r^{-D}(\delta_{ij} -D \,\hat r_i \hat r_j), \quad \hat r_i=r_i/|\vec r|\,,\label{eq:VD}
\end{equation}
The right-hand side of (\ref{eom}) describes a force induced by a dipole located at the origin with the strength $\dip$. $\vec a$ is a unit vector that describes the direction of the dipole. We treat it as a random variable, which makes (\ref{eom}) a stochastic differential equation. 

Eq.(\ref{eom}) was proposed in the companion paper \cite{AFKMMSSYY} as a toy model that captures some features of the fluid turbulence.
The authors performed a numerical simulation of the random process, revealing that the particle's trajectory has a fractal dimension $2.4\sim 2.7$ in $D=3$ and $1.7\sim 1.9$ in $D=2$. It implies that the particle motion is not the normal Brownian motion. 

It motivates us to study (\ref{eom}) analytically. In this paper, we first derive the Fokker-Planck (FP) equation for the random process using the path-integral formalism \cite{PI} (section 2). The radial part of the FP equation agrees with a special case of the O’Shaughnessy-Procaccia anomalous diffusion equation \cite{OP}.  We solve the Fokker-Planck equation by combining Bessel functions (section 3).\footnote{We note that the exact solution in \cite{OP} is a special one where the particle starts from the origin $r=0$. For our purse, we need solutions where the particle starts somewhere else $r\neq 0$. } It has an unusual feature that the total probability is not invariant due to the singularity at the origin. It corresponds to a phenomenon that the particle, which gets too close to the origin, jumps because of the strong force from the dipole. The trajectory resembles the L\'evy flight \cite{LevyFlight} and has a nontrivial fractal dimension in three dimensions. At the same time, some portion of the particles jumps beyond the cut-off, which causes the decrease of the total probability. In \cite{AFKMMSSYY}, to estimate the fractal dimension, the authors used two methods, (1) the periodic boundary condition to recover the lost particles and (2) reset the lost particle to the initial position and restart. These two approaches give similar results.\footnote{To be precise, the second approach gives slightly smaller fractal dimensions.} We use the second idea to modify the probability density such that the total probability is conserved.

Using the explicit form of the modified Green's function, we evaluate the fractal dimension of the streamline (section 4). We first show the loss of the probability appears in the early time stage and argue that the modification of Green's function to recover the unitarity. 
The second characteristic feature  (\ref{eom}) is the diffusion is anisotropic, which we explain by showing the plot and computing the Hurst exponents for the radial and angular direction independently. We compute the latent fractal dimension by combining these two. The results meet the numerical simulation in \cite{AFKMMSSYY}.

The process (\ref{eom}) is interesting not only in the context of \cite{AFKMMSSYY} but in more general mathematical physics.
The scaling behavior in the radial direction has some similarities with the conformal invariant process (see, for example, \cite{Bessel}), which is tightly related to Stochastic Loewner Evolution (SLE) \cite{SLE}. SLE describes the critical phenomena of many lattice models in two dimensions with fractal curves as the boundaries of their clusters (or domain walls) \cite{SLEREVIEW}. In particular, SLE has recently been discussed as a model for two-dimensional turbulence \cite{SLETUR}. We discuss the connection between our model and SLE and how our model can be regarded as a kind of generalization of SLE (in Appendix \ref{SLEBessel}).

While this paper is motivated by \cite{AFKMMSSYY}, we mainly focus on the derivation of the Fokker-Planck equation through a path integral and the analytic study of the solutions thus obtained. Since it involves technique which may not be popular among the community of the fractal physics, we decided publish the results in the separated papers.

\section{Derivation of Fokker-Planck equation}
We first note that the random variable $\vec a$ is constrained by $|\vec a|=1$, which makes the derivation of the Fokker-Planck equation more involved. We use the path integral to define the probability density to describe the constraint.

\subsection{Path integral approach}
We start by discretizing the time interval by $t_i=t_0+i \Delta t$ ($i=0,1,2,\cdots N$). As an initial condition, particle's location is set to $\vec r_0$ at $t=t_0$. For each time-step, we use a discretized version of (\ref{eq:VD}),
\begin{equation}\label{eq:VDd}
	\vec r(t_{i+1})= \vec r(t_i) + d_H \Delta t\sum_j V_{ij}^{(D)}a_j
\end{equation}
for the time evolution. Since $a_j$ is a random variable, we have to introduce a probability distribution $P(\vec r, t_i: \vec r_0, t_0)$ at each $t_i$, which gives the probability for the particle to be located in an arbitrary domain $D$ at $t_i$ is,
\begin{equation}
\int_D d^D r P(\vec r, t_i : \vec r_0, t_0 )\,.
\end{equation}
At $t=t_0$, it is given by the delta function,
\begin{equation}
	P(\vec r, t_0:\vec r_0, t_0)=\delta^{(D)}(\vec r-\vec r_0)\,.
\end{equation}
At $t=t_1=t_0+\Delta t$, Eq.(\ref{eq:VDd}) implies,
\begin{equation}
P(\vec r_1, t_1 : \vec r_0, t_0 )=\int \tilde d\vec a \, \delta^{(D)} \left(\frac{\vec r_1-\vec r_0}{\Delta t}- \dip\, V(\vec r_0) \cdot \vec a\right)\,,
\end{equation}
where $\int \tilde d\vec a$ is an integration over the unit $(D-1)$-sphere.
$V(\vec r)$ is a $D\times D$ matrix $V^{(D)}_{ij}$ defined in eq.(\ref{eq:VD}). We will omit the upper index $D$ when there is no confusion.

The probability distribution after two steps can be evaluated by composing two $P$'s,
\begin{equation}
	\int_V d^{D} r_2 P(\vec r_2, t_2: \vec r_0, t_0 )
\end{equation}
with
\begin{align}
	& P(\vec r_2, t_2:  \vec r_0, t_0 )\\
	&~~~~~=\int d^D r_1 P(\vec r_2, t_2: \vec r_1, t_1)\,P(\vec r_1: t_1: \vec r_0, t_0,)
\end{align}
One may repeat such process to the arbitrary steps $N$.
\begin{align}
	& P(\vec r_N, t_N : \vec r_0,  t_0)\\
	&~~~~~=\int \prod_{i=1}^{N-1} d^D r_i 
	\prod_{i=0}^{N-1} P(\vec r_{i+1}, t_0+(i+1)\Delta t : \vec r_{i}, t_0+i \Delta t)
\end{align}

We rewrite it in the form of the path integral. We first note that
\begin{align}
	&P(\vec r_1, t_0+\Delta t: \vec r_0, t_0)\\
	&=\int \tilde d\vec a \, \delta^{(D)} \left(\frac{\vec r_1-\vec r_0}{\Delta t}-
	\dip\,V(\vec r_0) \vec a\right)\\
	&=\int \tilde d\vec a \int d^D\vec p\, \exp\left(\mathrm{i}\vec p\cdot
	\left(\frac{\vec r_1-\vec r_0}{\Delta t}-  \dip\,V(\vec r_0) \vec a
	\right)\right)
\end{align}
It implies that the expression $P_N$ is written as,
\begin{align} 
	& P(\vec r_N, t_N\ :   \vec r_0, t_0)\\
	&~~~~~=\int \prod_{i=1}^{N-1} d^D \vec r_i \prod_{j=0}^{N-1} \tilde d\vec a_j \, d^D\vec p_j \exp\left(\mathrm{i}\vec p_j\cdot
	\left(\frac{\vec r_{j+1}-\vec r_j}{\Delta t}-\dip\,V(\vec r_j) \vec a_j \right)\right)
\end{align}
One may rewrite it in a formal notation of the path integral,
\begin{align}
	& P(\vec y, t_1 :   \vec x, t_0)=\int D\vec r \,\tilde D\vec a \,D\vec p\, \exp\left(\mathrm{i}\int_{t_0}^{t_1} dt \,\vec p(t)\cdot
	\left(\frac{d\vec r(t)}{d t}- \dip\,V(\vec r(t))\vec a\right)\right)\label{PI1}
\end{align}
with the initial conditions $\vec r(t_0)=\vec x$ and $\vec r(t_1)=\vec y$. $\tilde D\vec a$ is the path integral measure over the unit sphere.

\paragraph{Removing the constraint on $\vec a$:}
In the expression (\ref{PI1}), the integration over $\vec a$ is constrained by $|\vec a |=1$.  A standard way to remove it is to introduce an auxiliary variable $\lambda$ and rewrite the integration as,
\begin{equation}
	\int \tilde d\vec a = \int d\vec a \,\delta(\vec a\cdot \vec a-1)
	\propto \int d\vec a\,d\lambda\, \exp\left(\frac{\mathrm{i}\lambda}{2}
	(\vec a\cdot \vec a-1)\right)
\end{equation}
There is no constraint on the integration variable $\vec a$ in the final expression.

The path integral is rewritten as,
\begin{align}
	& P(\vec y, t_1 : \vec x,  t_0)\\
	&~~=\int D\vec r \,D\vec a \,D\vec p\,D\lambda\\
	&~~\cdot\exp\int_{t_0}^{t_1} dt\left(\frac{\mathrm{i}\lambda(t)}{2}(\vec a(t)\cdot \vec a(t)-1)+
	\mathrm{i}\vec p(t)\cdot
	\left(\frac{d\vec r(t)}{d t}-\dip\,V(\vec r(t))\cdot \vec a(t)\right)\right)\,.
	\label{PI2}
\end{align}

\paragraph{Gaussian integration over various variables:}
We note that the integrand in (\ref{PI2}) is gaussian with respect to $\vec a$.
After the integration over $\vec a$, we arrive at,
\begin{align}
	& P(\vec y, t_1 : \vec x,  t_0 )=\int D\vec r D\vec p D\lambda\, e^{\mathrm{i} S_1(\vec r, \vec p, \lambda)}\\
	& S_1(\vec r, \vec p, \lambda)=\int dt\left(-\frac{\lambda}{2} -\frac{\dip^2}{2\lambda}\vec p \cdot V(\vec r)^2\cdot \vec p +\vec p\cdot \frac{d\vec r}{dt}\right)\,.
\end{align}
$S_1$ is again quadratic with respect to $\vec p$. After the Gaussian integration over $\vec p$, one arrives at,
\begin{align}
	& P(\vec y, t_1:   \vec x, t_0)=\int D\vec r D\lambda \,  e^{\mathrm{i} S_2(\vec r, \lambda)}\label{PI3}\\
	&S_2(\vec r, \lambda)=\int dt \,\frac{\lambda(t)}{2}\left(\dip^{-2}\frac{d\vec r}{dt} V(\vec r)^{-2} \frac{d\vec r}{dt}-1\right)\label{S2}
\end{align}
We will use this expression to derive the Fokker-Planck equation.

\subsection{Derivation of Fokker-Planck equation} 
The analysis of the path integral of the form (\ref{S2}) was considered in the literature (see \cite{Polyakov} for instance)when the background is flat $V_{ij}=\delta_{ij}$. When we divide the time interval into large enough numbers, one may replace the dynamical variable $\lambda$ to be a constant $\lambda_0$.\footnote{One may apply the conventional quantization method of the constrained system. We give the outline in appendix \ref{DiracQ}.} The Lagrangian of the system may be written as,
\begin{equation}
	L=\frac{\lambda_0}2\left(\dip^{-2}\frac{d r^i}{dt} (V^{-2})_{ij}(\vec r) \frac{d r^j}{dt}-1\right)
\end{equation}
The Hamiltonian which comes from this Lagrangian is,
\begin{equation}
\mathcal{H}=\frac{\dip^2}{2\lambda_0}
 p_i (V^2)_{ij}  p_j+\frac{\lambda_0}{2}\,,\quad p_i=\lambda_0 \dip^{-2}(V^{-2})_{ij}\frac{dr^i}{dt}\,.
\end{equation}
We note that,
\begin{equation}
(V^2)_{ij}=r^{-2D}(\delta_{ij}+(D^2-2D)\hat r_i \hat r_j)=r^{-2D}((D-1)^2 \hat r_i \hat r_j +\sum_\alpha^{D-1} \hat e_\alpha \hat e_\alpha)\,.
\end{equation}

where $\hat e_\alpha$ is the orthonormal basis in the angular direction.
We replace $p_i\to \mathrm{i}\partial_i$ in the orthogonal coordinate. In the polar coordinates in $D$ dimensions, one may write\footnote{As it is obvious in the formula, $D=1$ is singular since $r$ derivatives vanish. We restrict ourselves to $D>1$ in the following.
},
\begin{equation}\label{Hamil}
\mathcal{H}=-\frac{\dip^2}{2\lambda_0 r^{2D+2}}\left(
 (D-1)^2 (r^2\partial_r^2-(D+1)r\partial_r)+\hat\Omega \right)+\frac{\lambda_0}{2}.
\end{equation}
Here $\hat\Omega$ is the Laplacian on $S^{D-1}$ written in terms of the angular variables. For instance,
\begin{equation}
\hat\Omega=\left\{
\begin{array}{ll}
\partial_\theta^2\quad& \mbox{for }D=2\\
\frac{1}{\sin\theta}\partial_\theta (\sin\theta\partial_\theta)+\frac{1}{\sin^2\theta}\partial_\varphi^2\quad & \mbox{for } D=3
\end{array}
\right.
\end{equation}
We note some ambiguity in the ordering of $r$ and $\partial_r$. To fix it, we require
the radial Hamiltonian is Hermitian and total derivative with respect to the measure $r^{D-1}dr$:
\begin{equation}
 (r^{D-1}\Delta_r)^\dagger = r^{D-1}\Delta_r, \quad r^{D-1}\Delta_r=\partial_r(\mbox{something}),\quad 
\end{equation}
for $\Delta_r=r^{-2D-2}(r^2\partial_r +a r\partial_r +b)$ with constant $a,b$. These conditions give $a=-(D+1)$ and $b=0$.

The path integral formula implies that the probability density $P(\vec r, t)$ satisfies the following Fokker-Planck equation,
\begin{align}
\frac{\partial}{\partial t}P(\vec r,t)&=\frac{h}{2 r^{2D+2}}\left(
 (D-1)^2 (r^2\partial_r^2-(D+1)r\partial_r)+\hat\Omega \right)P-\frac{\lambda_0}{2}P,
 \label{FPD}
\end{align}
with $h=\dip^2/\lambda_0$\,. The last term in (\ref{FPD}) breaks the conservation of probability. Since it is arbitrary, we will use a tuning $\lambda_0\to 0$ while keeping $h$ finite. As we see, the total probability obtained by the integration of $P(\vec r, t)$ is not conserved due to the singularity at the origin $r=0$, where the last term may play some role in the future. For the following analysis, we will ignore it.

We note that the radial part of (\ref{FPD}) is an example of the anomalous diffusion equation obtained by O'Shaughnessy and Procaccia \cite{OP}.
The authors of \cite{OP} found an exact solution in a closed form,
\begin{equation}\label{solOP}
P(r,t)\propto t^{-\frac{D}{2+\theta}} \exp\left(-\frac{r^{2+\theta}}{K(2+\theta)^2 t}\right)\,,
\end{equation}
where $\theta$ is an arbitrary parameter of the equation. In our case, one should identify $\theta=2D$ and set $K=h/2$.
It describes an anomalous diffusion starting from $r=0$. In our case \cite{AFKMMSSYY}, we set the initial location of the particle outside of the origin. It requires us to study the more general families of the solutions, which reveals a new essential feature of the equation.

\paragraph{Scaling symmetry:}
A basic property of the Fokker-Planck equation is the existence of the scaling symmetry, (with $\bomega$, the angular coordinates),
\begin{equation}\label{scaling}
r\rightarrow r'= \gamma r,\quad \vec\omega\rightarrow \vec\omega,\quad t\rightarrow \gamma^{2D+2} t\,,
\end{equation}
for arbitrary $\gamma\in \mathbb{R}_{\neq 0}$.
It implies that the solution of (\ref{FPD}) with the initial value, say $P(t=0)\sim \delta(r-r_0)$ can be obtained from the initial value problem with $\delta(r-1)$ by scaling the time parameter $t\rightarrow  r_0^{2(D+1)}t$.

\paragraph{Boundary condition at $r=0$}
We normalize the probablity density $P(r,\bomega, t)$  by,
\begin{equation}
\int_0^\infty r^{D-1}dr \int_{S^{D-1}} d\bomega \,P(r,\bomega,t)=1\,.
\end{equation}
$d\bomega$ is a normalized measure on $S^{D-1}$. 
By taking the time derivative, one obtains,
\begin{align}
0&=\int_0^\infty r^{D-1}dr \int d\bomega \,\partial_tP(r,\bomega,t)\nonumber\\
&=\int_0^\infty dr \int d\bomega \,\frac{h}{2} \left((D-1)^2\partial_r \left(r^{-D-1}\partial_r P\right)+r^{-D-3}\hat\Omega P\right)-\frac{\lambda_0}{2}\nonumber\\
&=\left.\int d\bomega\,\frac{h(D-1)^2}{2}r^{-D-1}\partial_r P(r,\bomega,t)\right|_{r=0}-\frac{\lambda_0}{2}\,.\label{conservation}
\end{align}
Naively, it would be best to use the boundary conditions at $r=0$ and $r=\infty$,
\begin{equation}
\left. r^{-D-1}\partial_r P(r,\bomega,t)\right|_{r=0,\infty}=0\,.\label{BCD}
\end{equation}
In particular, the particular solution (\ref{solOP}) satisfies this condition.

As we will see, it is impossible to obtain a complete set of eigenfunctions which satisfies this boundary condition. One may interpret it as the breakdown of the unitarity of the Hamiltonian through the boundary condition, namely due to the singularity at $r=0$\footnote{As a different approach, a cut-off by $r=\epsilon$ is discussed in Section 4.4}.

In the exact solutions obtained in the next section, it is necessary to impose a relaxed condition at $r=0$,
\begin{equation}\label{BCr}
\lim_{r\to 0}  r^{-D-1}\partial_r P(r,\bomega,t)=\mbox{finite},
\end{equation}
to have nontrivial solutions. It implies that the probability flows out at the origin. As explained in the introduction, we interpret it that the particles, which approach too close to the origin, jump out of the cut-off as shown in the numerical simulation \cite{AFKMMSSYY}. The recovery of the probability depends on the numerical setup. We will study more detail in the next section.

\section{Analysis of Fokker-Planck equation}\label{s:Fractal}
\subsection{Exact solution in arbitrary dimensions}\label{s:DdimSOl}
It is straightforward to solve (\ref{FPD}) with arbitrary initial and the relaxed boundary conditions. We drop the last term in (\ref{FPD}).
We first construct the eigenfunctions of the hamiltonian $\mathcal{K}=\frac{2}{h}\mathcal{H}$,
\begin{equation}
\mathcal{K}\psi(r,\bomega)=-k^2 \psi(r,\bomega),\quad
\mathcal{K}=r^{-2D-2}\left(
 (D-1)^2 (r^2\partial_r^2-(D+1)r\partial_r)+\hat\Omega \right),
\end{equation}
where $k^2\geq 0$ is the eigenvalue. 
We use the standard separation of variable technique for the radial and a set of the angle variables $\vec\omega=(\omega_1,\cdots, \omega_{D-1})$,
\begin{align}
\psi(r,\omega)=R(r)\Theta(\bomega), \quad \hat\Omega \Theta(\bomega)=-b^2 \Theta(\bomega),
\end{align}
where $b^2\geq 0$ is the eigenvalue of $\hat{\Omega}$.
The eigenfunctions in the angular directions are,
\begin{align}
D=2:\quad&\Theta(\theta)=e^{\mathrm{i}n\theta} , \quad b ^2=n^2, \quad n\in \mathbb{Z}\\
D=3:\quad & \Theta(\theta,\varphi)=Y_{\ell m}(\theta,\varphi),\quad b^2=\ell(\ell+1) \quad \ell\in \mathbb{Z}_{\geq 0},\, m\in[-\ell,\ell]\,,
\end{align}
where $Y_{\ell,m}$ is the spherical harmonics.

The equation for $R(r)$ becomes,
\begin{equation}\label{RDE}
r^{-2D-2}((D-1)^2(r^2\partial_r^2-(D+1)r\partial_r)-b^2)R(r)=-k^2R(r)\,.
\end{equation}
We change variable 
\begin{equation}
\zeta=\frac{kr^{D+1}}{D^2-1}
\end{equation}
and rewrite $R(r)=r^{1+D/2} f(\zeta)$. 
The differential equation (\ref{RDE}) becomes the Bessel equation,
\begin{equation}
\zeta^2\frac{d^2 f}{d\zeta^2}+\zeta\frac{d f}{d\zeta}+(\zeta^2-\nu^2)f=0\,,
\end{equation}
with
\begin{equation}\label{nu}
	\nu=\frac{\sqrt{4b^2+((2+D)(D-1))^2}}{2(D^2-1)}.
\end{equation}
It implies that the general solutions are,
\begin{equation}\label{Rr}
R(r) = r^{1+D/2} \left(c_+ J_\nu\left(\frac{kr^{D+1}}{D^2-1}\right)+c_- J_{-\nu}\left(\frac{kr^{D+1}}{D^2-1}\right)\right)
\end{equation}
where $c_\pm$ are arbitrary constants.

In the vicinity $x\sim 0$, the Bessel function behaves as,
\begin{equation}
J_\nu(x)\sim \frac{1}{2^\nu \Gamma(\nu+1)}x^\nu(1+O(x^2))\,.
\end{equation}
One note that the solution $J_{-\nu}$ does not meet the boundary condition (\ref{BCD}) nor the relaxed one (\ref{BCr}). For $J_\nu$, 
\begin{equation}
r^{1+\frac{D}{2}} J_\nu\left(\frac{kr^{D+1}}{D^2-1}\right)\propto r^{1+\frac{D}{2}+\sqrt{\left(\frac{b}{D-1}\right)^2+(1+\frac{D}{2})^2}}(1+O(r^{2(D+1)}))\,.
\end{equation}
While $b^2>0$, the stronger condition (\ref{BCD}) holds. For the spherically symmetric case $b^2=0$, on the other hand, we can impose only the relaxed condition (\ref{BCr}). Since we need the spherically symmetric functions to form the complete basis, the violation of the boundary condition (\ref{BCD}) is inevitable.

From the complete set of solutions of (\ref{FPD}), one may write the general solutions as,
\begin{equation}
P(r,\vec\omega,t)=\int_0^\infty dk \sum_{\vec n} \rho_{\vec n}(k) 
e^{-\frac{htk^2}{2} } 
Y_{\vec n}(\vec\omega)
\left(\frac{kr^{D+1}}{D^2-1}\right)^{\frac{1+D/2}{D+1}} J_{\nu_{\vec n}}\left(\frac{kr^{D+1}}{D^2-1}\right),
\end{equation}
where $\vec\omega$ is the coordinate of $S^{D-1}$, $Y_{\vec n}(\vec\omega)$ is the spherical harmonics with the quantum numbers $\vec n$, $\nu_{\vec n}$ is (\ref{nu}) where $b$ is determined from $\vec n$.

\subsection{Green's function}
Green's function $G(r,\vec\omega,t)$ is the probability distribution $P(r,\vec \omega, t)$ that corresponds to the choice of the delta function distribution as the initial value. The support of the delta function is located at $r>0$.

The derivation of Green's function is elementary, by using the orthogonality of the spherical harmonics and the Hankel transformation for the Bessel functions,
\begin{equation}
	f(x)=\int_0^\infty kdk  \tilde f(k)J_\nu(kx),\qquad
	\tilde f(k)= \int_0^\infty xdx  f(x)J_\nu(kx)\,.\label{Hankel}
\end{equation}
In particular, we will use the following cases.
\begin{enumerate}
\item Spherically symmetric case for arbitrary $D$. We set the initial value as,
\begin{equation}
G(r,t=0)= \delta(r-1)\,.
\end{equation}
Green's function is,
\begin{equation}\label{GreenSpherical}
G(r,t)=\frac{r^{1+\frac{D}{2}}}{(D-1)^2(D+1)}\int_0^\infty kdk e^{-hk^2t/2}
J_{\nu_D}\left(\frac{k}{D^2-1}\right)J_{\nu_D}\left(\frac{kr^{D+1}}{D^2-1}\right)
\end{equation}
with 
\begin{equation}
\nu_D=\frac{1+D/2}{D+1}.\label{nuD}
\end{equation}
\item $D=2$ with the angle dependence. The initial value is
$G(r,\theta,t)=\delta(r)\delta(\theta)$. 
\begin{equation}\label{Green2D}
	G(r,\theta,t)= \frac{r^{2}}{6\pi}\int_0^\infty kdk\sum_{n\in \mathbb{Z}} J_{\nu_n}(kr^3/3)J_{\nu_n}(k/3)e^{-hk^2 t/2+\mathrm{i}n\theta}\,.
\end{equation}
\item $D=3$ with the angle dependence. With the polar coordinates $r, \theta,\varphi$, ($0\leq \theta\leq \pi$, $\varphi\in [-\pi, \pi]$), we set the initial condition,
\begin{equation}
G(r,\theta,\varphi,t)=\delta(r-1)\delta(\theta)\,.
\end{equation}
Green's function becomes,
\begin{equation}\label{green3D}
	G(r,\theta,\varphi,t)= \frac{r^{5/2}}{8}\int_0^\infty kdk \sum_{l=0}^\infty \frac{2l+1}{4\pi} J_{\nu_l}(kr^4/8)J_{\nu_l}(k/8)P_l(\cos\theta) e^{-hk^2 t/2}\,,
\end{equation}
where $P_l$ is the Legendre function.
\end{enumerate}
In appendix \ref{s:Green}, we give the explicit derivation of the formulae.

\subsection{Recovery of the unitarity through boundary conditions}
The conservation law
\begin{equation}\label{unitarity}
\int r^{D-1}drd\vec\omega P(r,\vec\omega,t) =1
\end{equation}
is necessary to interpret $P(r,\vec\omega,t)$ as the probability distribution. This condition, however, can be broken through the use of the relaxed boundary condition (\ref{BCr}) for the spherically symmetric function. It applies to Green's function obtained in the last subsection.

When the unitarity (\ref{unitarity}) is broken, we have to modify the function $P(r,\vec\omega,t)$ to define the diffusion process properly. It depends on the set-up of the numerical simulation, with which we compare the result. In our case, we need a comparison with the companion paper \cite{AFKMMSSYY}, where the authors consider the motion in the space in a square (or a cube) with the edge length $2L$. When the particle jumps out of the box because of the vital force in the vicinity of the dipole, they use two types of ``boundary conditions" to recover the particles.
\begin{itemize}
\item Condition 1 uses the periodic boundary condition to force the particle to return to the original square or cube.
\item Condition 2 forces the disappeared particle to the initial position of the particle.
\end{itemize}
Under these boundary conditions, they obtained the fractional dimension of the trajectory $D_f$ of the particle. For $D=3$ with condition 1, $D_f\sim 2.6$ and with condition 2, $D_f\sim 2.4$. For $D=2$ with condition 1, $D_f\sim 1.8$, and with the condition 2, $D_f\sim 1.7$. 

We modify Green's function for the condition 2. We define the total probability from the original Green's function as,
\begin{equation}
N(t)=\int r^{D-1}drd\vec\omega P(r,\vec\omega,t)\,.
\end{equation}
One can evaluate the derivative with respect to time by (\ref{conservation}).

From the solution of the FP equation, $P(r,\vec\omega,t)$, we define the modified probability density $\tilde P$ by using the condition 2,
\begin{align}
\tilde P(r,\vec\omega,t) &= P(r,\vec\omega,t)-\int_0^t dt_1 \dot N(t_1) P(r,\vec\omega,t-t_1)\nonumber\\
&+ \int_0^t dt_1 \dot N(t_1)\int_{t_1}^t \dot N(t_2) P(r,\vec\omega,t-t_2)+\cdots\,,
\label{Q1}\\
&= P(r,\vec\omega, t)-\int_0^t dt_1 \dot N(t_1) \tilde P(r,\vec\omega, t-t_1)\,,\label{Q2}
\end{align}
where $\dot N=\frac{dN}{dt}$. 
The second term in the first line implies the recovery of the particle during the period $[0,t]$. The second line of (\ref{Q1}) is necessary that the particle, which was once recovered, can disappear again during the time interval $[t_1,t]$. Eq.(\ref{Q2}) gives a recursive definition of $\tilde P$. $\tilde P$ satisfies
$\int r^{D-1}drd\vec\omega \tilde P(r,\vec\omega,t)=1$. \footnote{To prove it, one integrate (\ref{Q1}) over the space. One obtains an integral equation, $\tilde N(t)=N(t)-\int_0^{t} dt_1 \dot N(t_1) \tilde N(t-t_1)$ where $\tilde N(t)=\int r^{D-1}drd\vec\omega Q(r,\vec\omega,t)$. An obvious solution is $\tilde N(t)=1$, but it is also the unique solution due to the nature of the integral equation.} Thus, one may use it as the definition of the probability density.

\section{Numerical analysis and the fractal diffusion}
While (\ref{GreenSpherical}--\ref{green3D}, \ref{Q2}) gives the explicit form of (modified) Green's function, it is not so obvious to find the fractal behavior of the trajectory in an analytic form since it involves the integration and the summation. In the following, we evaluate these expressions numerically by cutting off the $k$ integration to $k\in (0,k^\mathrm{max})$ and the summation over the angular labels $|n|<n^\mathrm{max}$ at $k^\mathrm{max}\sim 60$ and $n^\mathrm{max}\sim 50$ respectively, and set $h=1$. $k^\mathrm{max}$ and $n^\mathrm{max}$ give the ultra-violet cut-off, which is necessary for numerical computation. If we choose them too small, Green's functions have un-physical oscillating behavior, making the fractal dimension calculation ill-defined. As they become larger, one has a better description for smaller t.We use {\sl Mathematica} for the evaluation.

In this section, we show
\begin{itemize}
\item The decrease of the total probability due to the singularity at the $r=0$. It appears relatively early, $t\sim 0.01$ for $D=2,3$. We need to use the modified probability density (\ref{Q2}) to obtain the proper result.
\item The diffusion is anisotropic.
\item The latent fractal dimensions, in the sense of Mandelbrot, coincides with the numerical results in \cite{AFKMMSSYY}
\end{itemize}

\subsection{Decrease of the total probability}
So far, we have discussed the loss of the total probability by analytical methods.
To illuminate how it appears, we plot Green's function for the spherically symmetric case (\ref{GreenSpherical}) for $D=2, 3$.
 
 \begin{figure}[btp]
    \begin{tabular}{cc}
      \begin{minipage}[t]{0.45\hsize}
        \centering
        \includegraphics[scale=0.5]{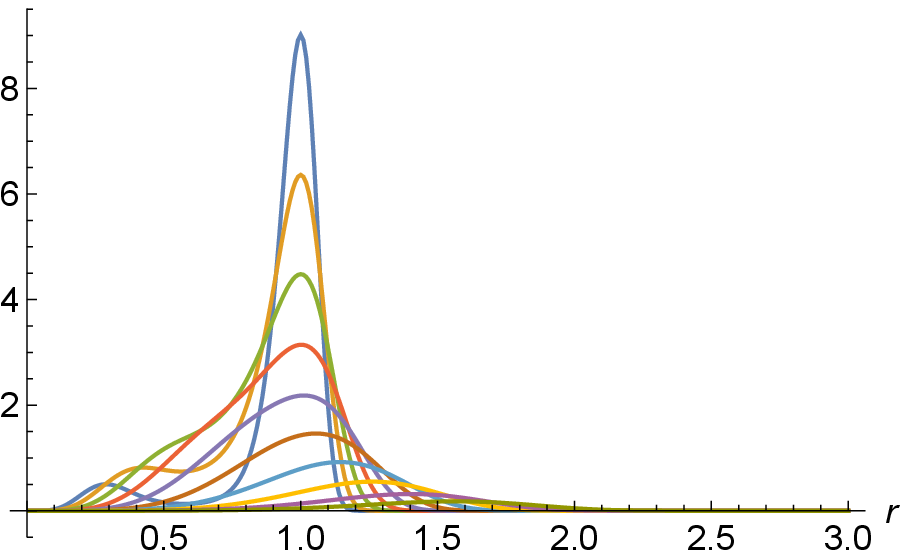}
        \subcaption{$D$=2}
        \label{radial_pD2}
      \end{minipage} &
      \begin{minipage}[t]{0.45\hsize}
        \centering
        \includegraphics[scale=0.5]{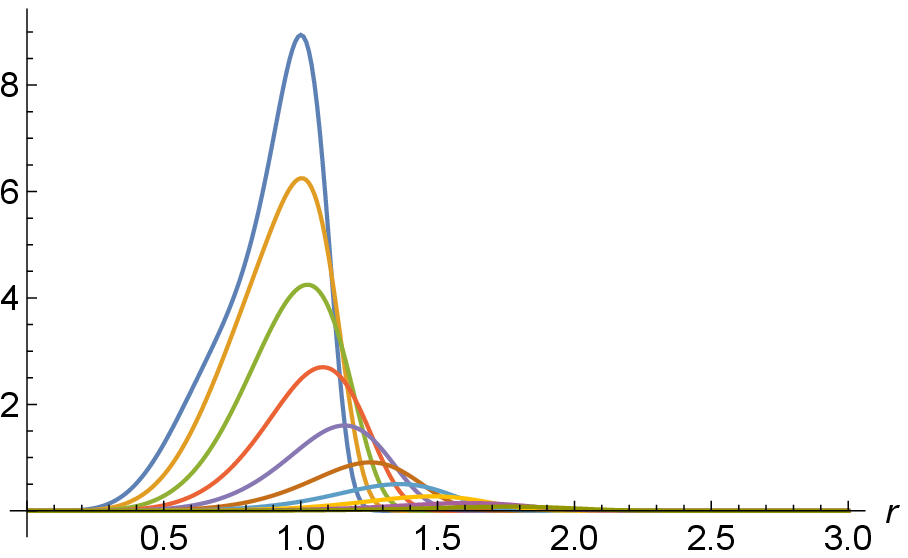}
        \subcaption{$D=3$}
        \label{radial_pD3}
      \end{minipage} 
    \end{tabular}
     \caption{Plots of $G(r,t)$ for $t=2^n$ ($n=-9,-8,\cdots, 0$). Horizontal axis gives the radius}
     \label{fig:RadialProfile}
  \end{figure}
  
  In Figure \ref{fig:RadialProfile}, we plot $P(r,t)$ for $D=2,3$ with $t=2^{n}$ ($n=-9,-8,\cdots,0$) with different colors for each $t$. The sharpest profile corresponds to $t=2^{-9}$, and the curve blurs as the time evolution. The total probability equals the area sandwiched between the plot curve and the horizontal axis. It begins to decrease the time evolution in the early stage, say $t=2^{-7}\sim 0.01$, and most of the probability vanishes as early as $t\sim 0.25$. We also note that the shape of the plot implies that the particles are absorbed to the origin, which we interpret as the repulsion to infinity. If one compares $D=2$ and $D=3$, the repulsion at the origin is stronger for $D=3$ since the singularity at $r=0$ is stronger for $D=3$.



\begin{figure}[btp]
    \begin{tabular}{cc}
      \begin{minipage}[t]{0.45\hsize}
        \centering
        \includegraphics[scale=0.5]{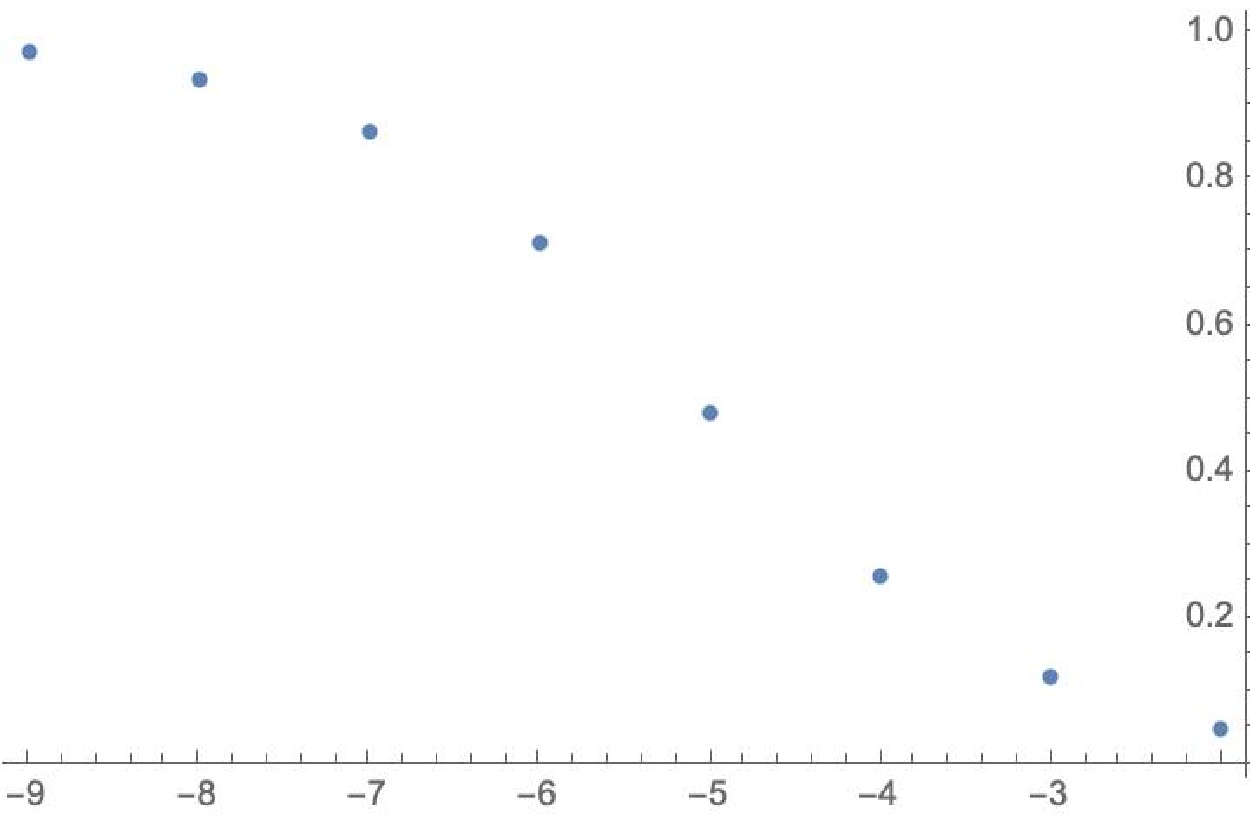}
        \subcaption{$D=2$}
        \label{TP2D}
      \end{minipage} &
      \begin{minipage}[t]{0.45\hsize}
        \centering
        \includegraphics[scale=0.5]{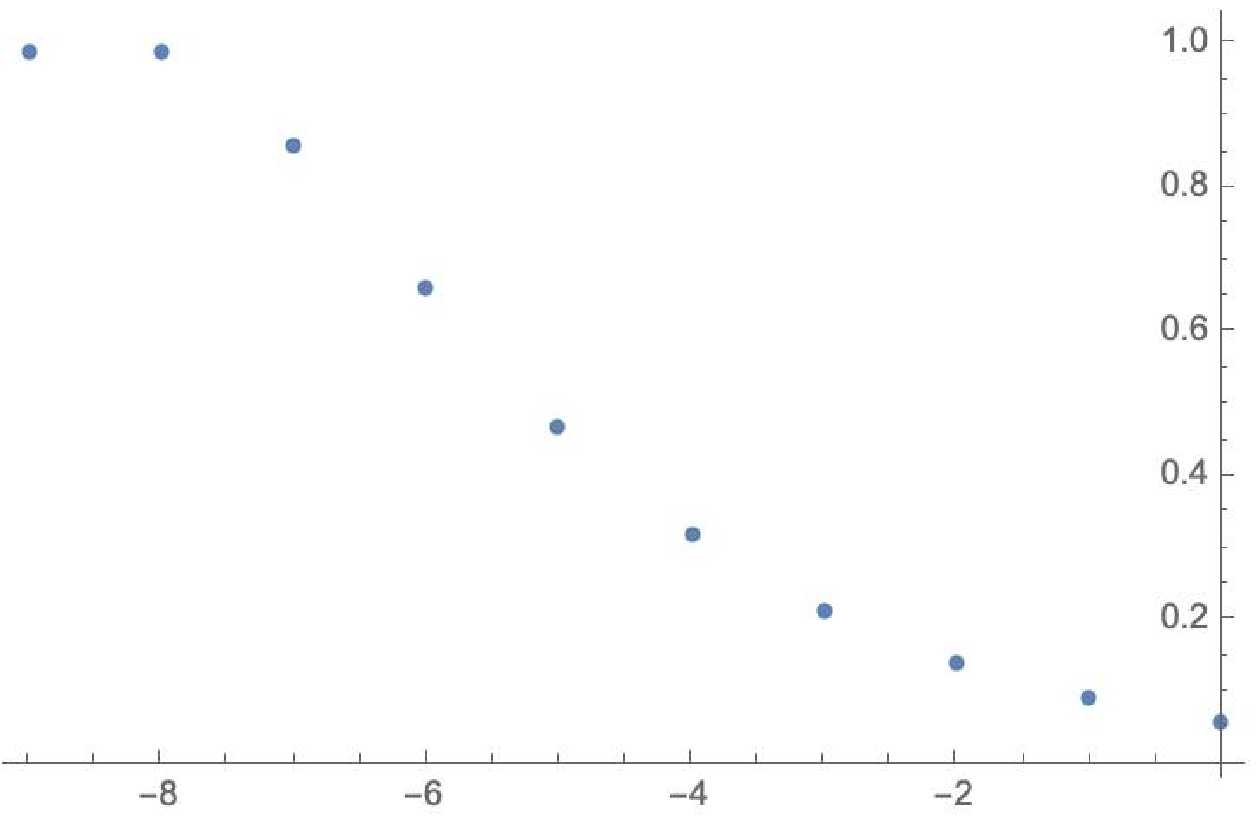}
        \subcaption{$D=3$}
        \label{TP3D}
      \end{minipage} 
    \end{tabular}
     \caption{Decrease of total probability. Horizontal axis gives $\log_2 t$.}
     \label{fig:TP}
  \end{figure}

The absorption of the total probability remains the same if we include the angular variable dependence. We compute the total probability, which depends on time for $D=2, 3$ in Figure \ref{fig:TP}. The horizontal axis describes the time ($\log_2 t$), and the vertical axis gives the total probability. Both graphs describe a similar process with a slight shift in the decreasing time interval; the absorption is slower in $D=2$. In either case, one observes the total probability vanishes in a relatively small time interval $t\sim 0.05$. We need the modification of Green's function to evaluate the fractal dimensions.
We note that there are similar plots in \cite{AFKMMSSYY}.

 
 \subsection{Profile of modified Green's function}
\begin{figure}[t]
    \begin{tabular}{cc}
      \begin{minipage}[t]{0.45\hsize}
        \centering
        \includegraphics[keepaspectratio, scale=0.3]{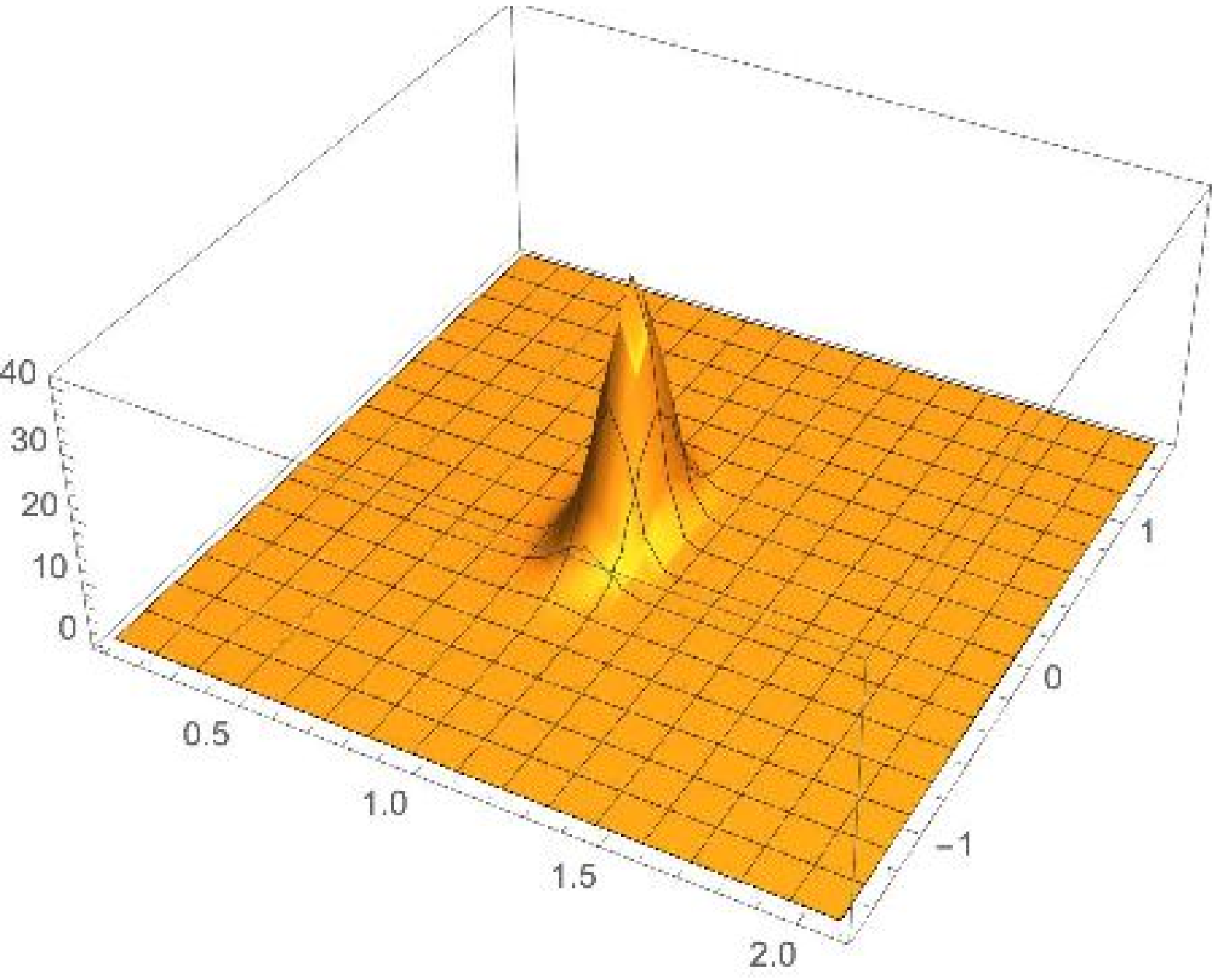}
        \subcaption{$t=2^{-8}$}
        \label{t2-8}
      \end{minipage} &
      \begin{minipage}[t]{0.45\hsize}
        \centering
        \includegraphics[keepaspectratio, scale=0.3]{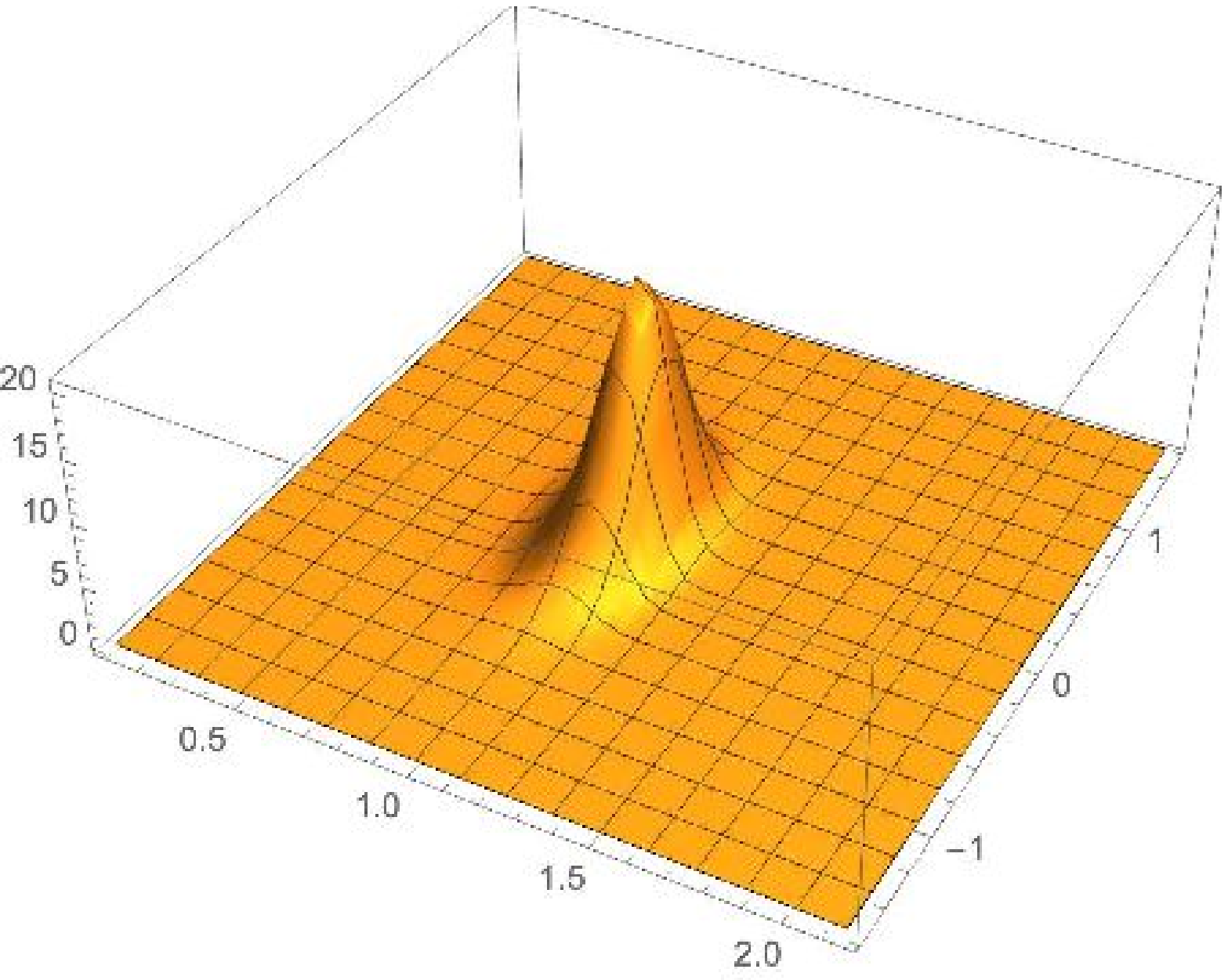}
        \subcaption{$t=2^{-7}$}
        \label{t2-7}
      \end{minipage} \\
   
      \begin{minipage}[t]{0.45\hsize}
        \centering
        \includegraphics[keepaspectratio, scale=0.3]{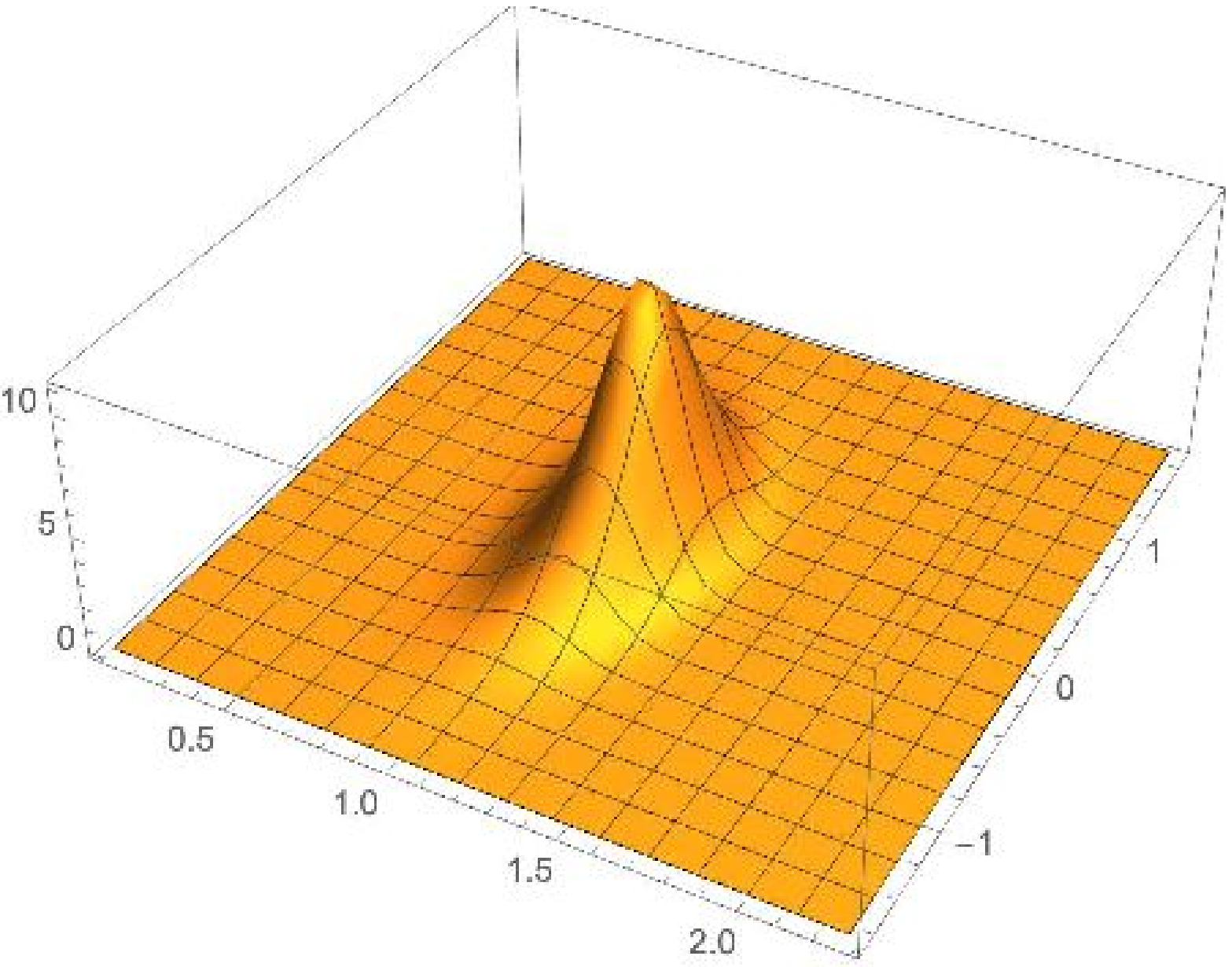}
        \subcaption{$t=2^{-6}$}
        \label{t2-6}
      \end{minipage} &
      \begin{minipage}[t]{0.45\hsize}
        \centering
        \includegraphics[keepaspectratio, scale=0.3]{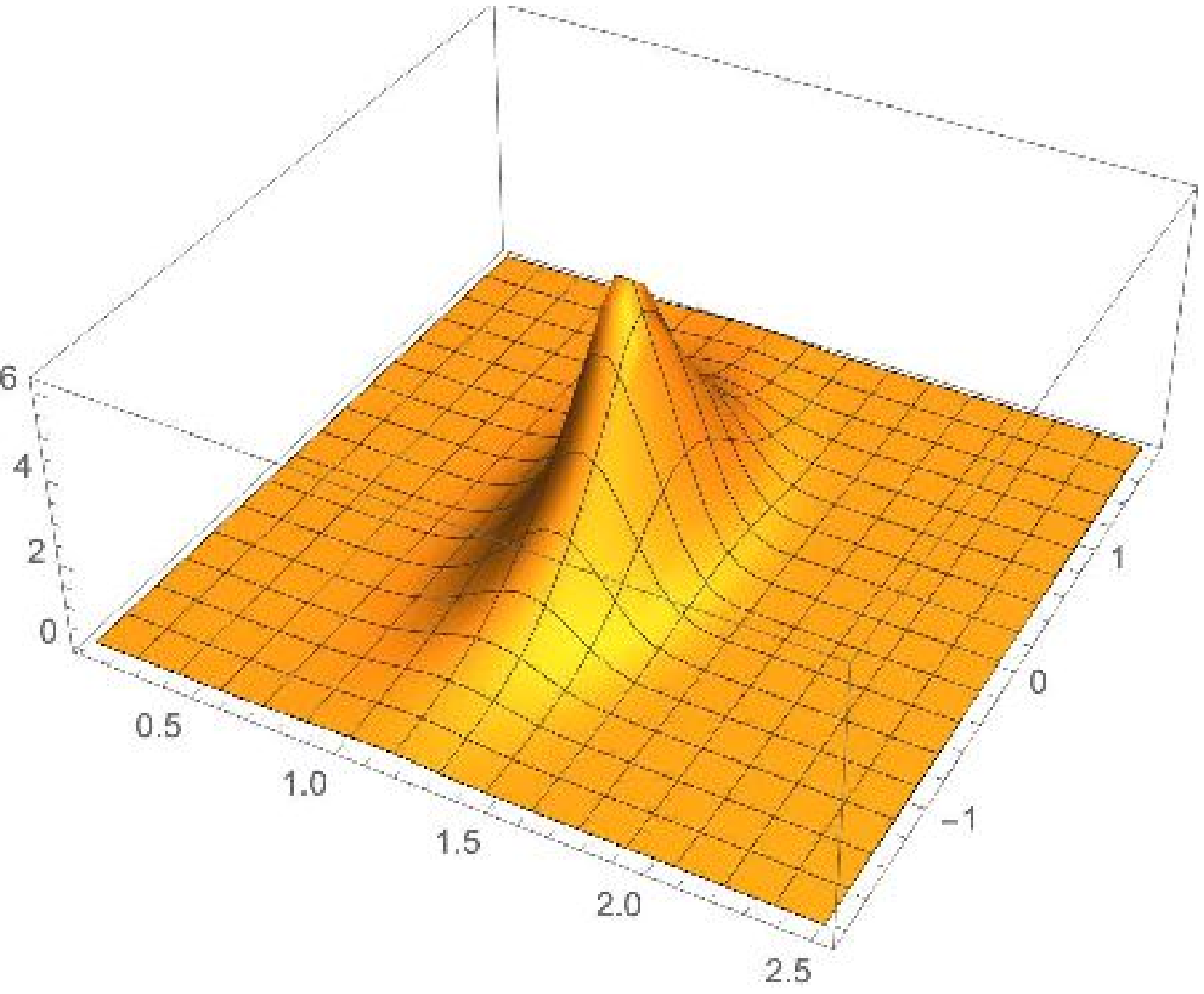}
        \subcaption{$t=2^{-5}$}
        \label{t2-5}
%
      \end{minipage} 
    \end{tabular}
     \caption{Time evolution of modified Green's function for $D=2$: The axis downward to right gives the radius $r\in [0.5,3]$ and the axis upward to right gives the angle $\theta\in[-\frac{\pi}{2}, \frac{\pi}{2}]$.
	 }
     \label{fig:Green2D}
  \end{figure}

The second characteristic feature of the system is that the behaviors in the radial direction and the angular directions are generally different. 
To see it graphically, we give plot of the modified Green's function for $D=2$ with $t=2^{n}$ with $n=-8,-7,-6,-5$ in Figure \ref{fig:Green2D}.

One may see that the diffusion in the angular direction is faster than the radial direction. It comes from the behavior of the original Green's function. At the same time, the shape of the sharp peak remains the same, which is due to the modification of Green's function (\ref{Q1}), where we have to add Green's function at an earlier time to recover the unitarity. We obtained a similar behavior for $D=3$; the diffusion in the radial direction is suppressed compared with the angular direction.

\subsection{Hurst exponent and Fractal dimension}
We compute the variation to evaluate the fractal dimension of the trajectory. Since the diffusion is anisotropic, separating the radial and the angular directions will be helpful.

We use the modified Green's function, $\tilde G$ (\ref{Green2D}, \ref{green3D}) with the modification (\ref{Q1}) to define the average. In the numerical computation, we truncate the summation to $\dot N^2$ order.

We use the polar coordinates to evaluate the variations.
For $D=2$, since we start from $\theta=0$, one may put $\langle\theta\rangle=0$ and $\langle \vec r\rangle=(\langle r\rangle,0)$. The variation is decomposed as,
\begin{align}
\langle (\vec r-\langle\vec r\rangle)^2\rangle &=\langle (\Delta \vec r)^2= \langle r^2\rangle+\langle r\rangle^2-2\langle r\rangle\langle r\cos\theta\rangle=(\Delta_r)^2+(\Delta_\theta)^2\,,\\
(\Delta_r)^2&=\langle r^2\rangle-\langle r\rangle^2\,,\quad
(\Delta_\theta)^2=2\langle r\rangle(\langle r\rangle-\langle r\cos\theta\rangle)\,.
\end{align}
We refer to $\Delta_r$ (resp. $\Delta_\theta$) as the variation in the radial (resp. angular) direction.

For $D=3$ case, Green's function is symmetric for the azimuthal angle $\varphi$. Also, since we start from the north pole, we may assume the average of the polar coordinate $\langle\theta\rangle=0$ and $\langle \vec r\rangle=(0,0,\langle r\rangle)$. We obtain the same expression for the variation and the decomposition into the radial and angular directions.

We introduce Hurst exponents as,
\begin{equation}
\langle (\Delta \vec r)^{2}\rangle \propto t^{2H},\quad
(\Delta_r)^2 \propto t^{2H_r},\quad 
(\Delta_\theta)^2 \propto t^{2H_\theta}\,.
\end{equation}
where $0<H, H_r, H_\theta<1$. $H=H_r=H_\theta=1/2$ corresponds to the normal Brownian motion. Mandelbrot's latent fractal dimension is given by $D_f=1/H$. We note that the definition of the fractal dimension in \cite{AFKMMSSYY} is based on the box counting, which is different from the one here. While they do not necessarily give the same answer, these are only quantities which can compared with.

\begin{figure}[btp]
    \begin{tabular}{cc}
      \begin{minipage}[t]{0.45\hsize}
        \centering
        \includegraphics[keepaspectratio, scale=0.4]{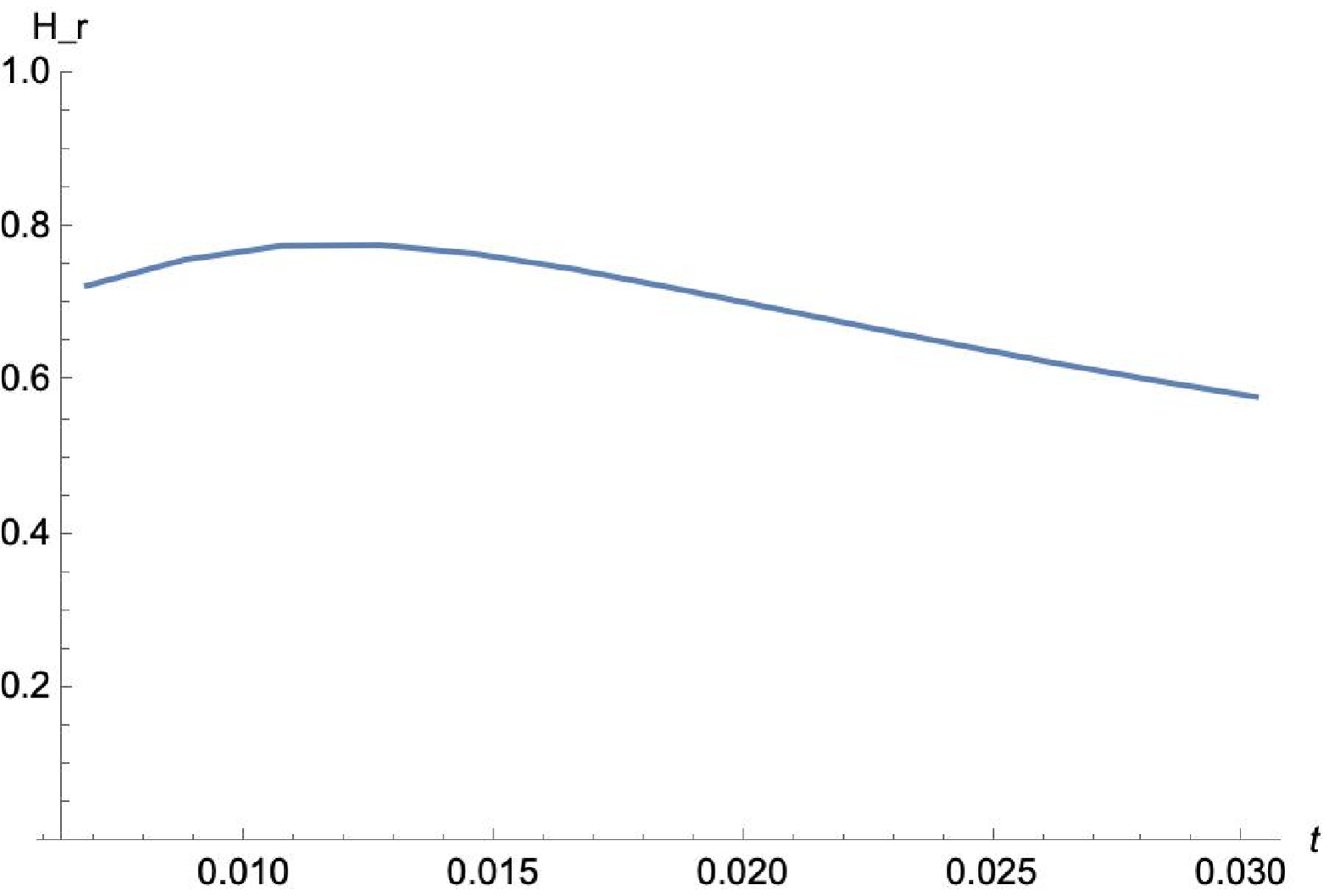}
        \subcaption{$H_r$}
        \label{2DHr}
      \end{minipage} \hfill&
      \begin{minipage}[t]{0.45\hsize}
        \centering
        \includegraphics[keepaspectratio, scale=0.4]{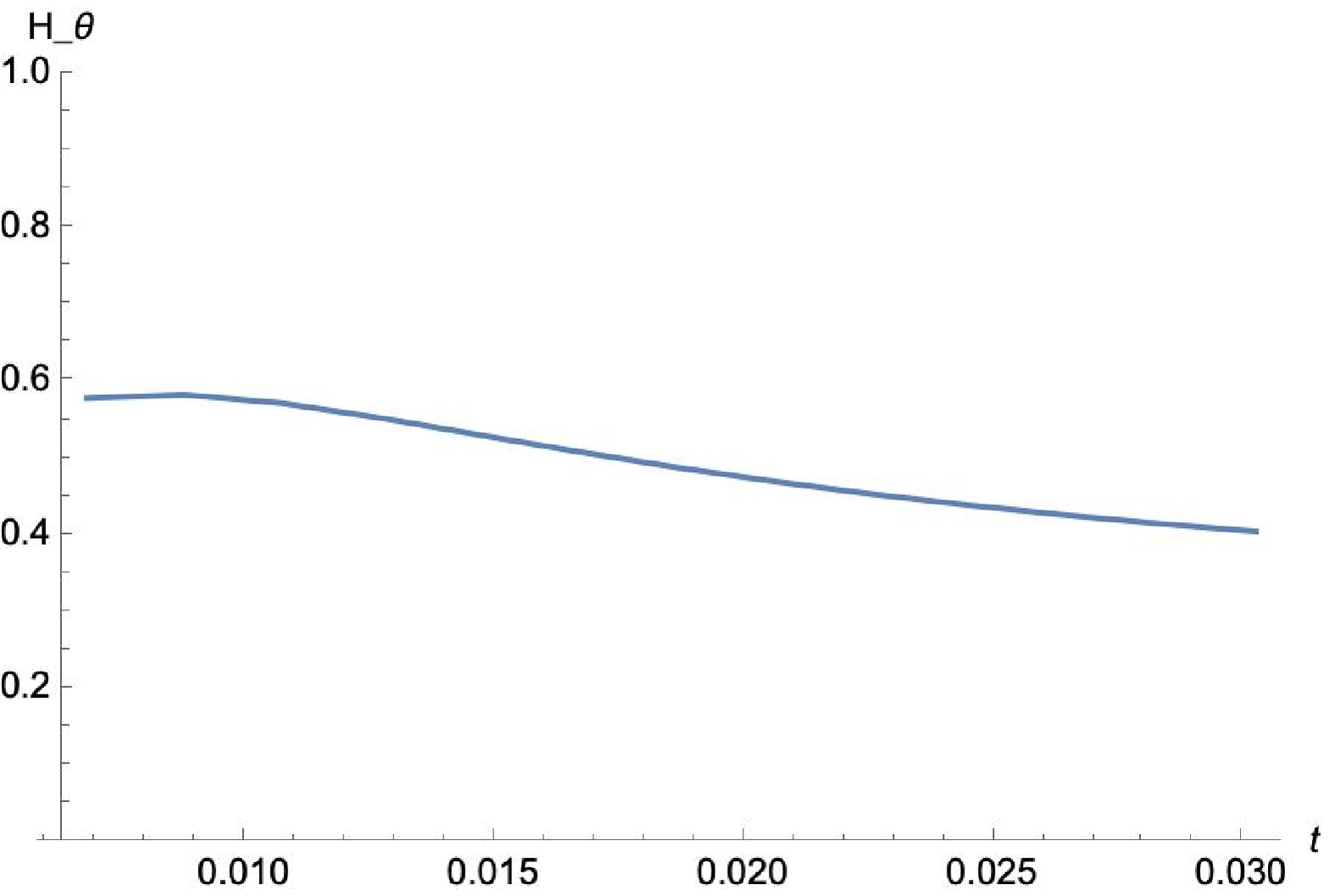}
        \subcaption{$H_\theta$}
        \label{2DHt}
      \end{minipage}
    \end{tabular}
     \caption{$H_r$ and $H_\theta$ for $D=2$. Horizontal axis describes time $t$.}
 \label{fig:D2H}
  \end{figure}

\begin{figure}[tb]
	\begin{tabular}{cc}
		\begin{minipage}[t]{0.45\hsize}
			\centering
			\includegraphics[keepaspectratio, scale=0.4]{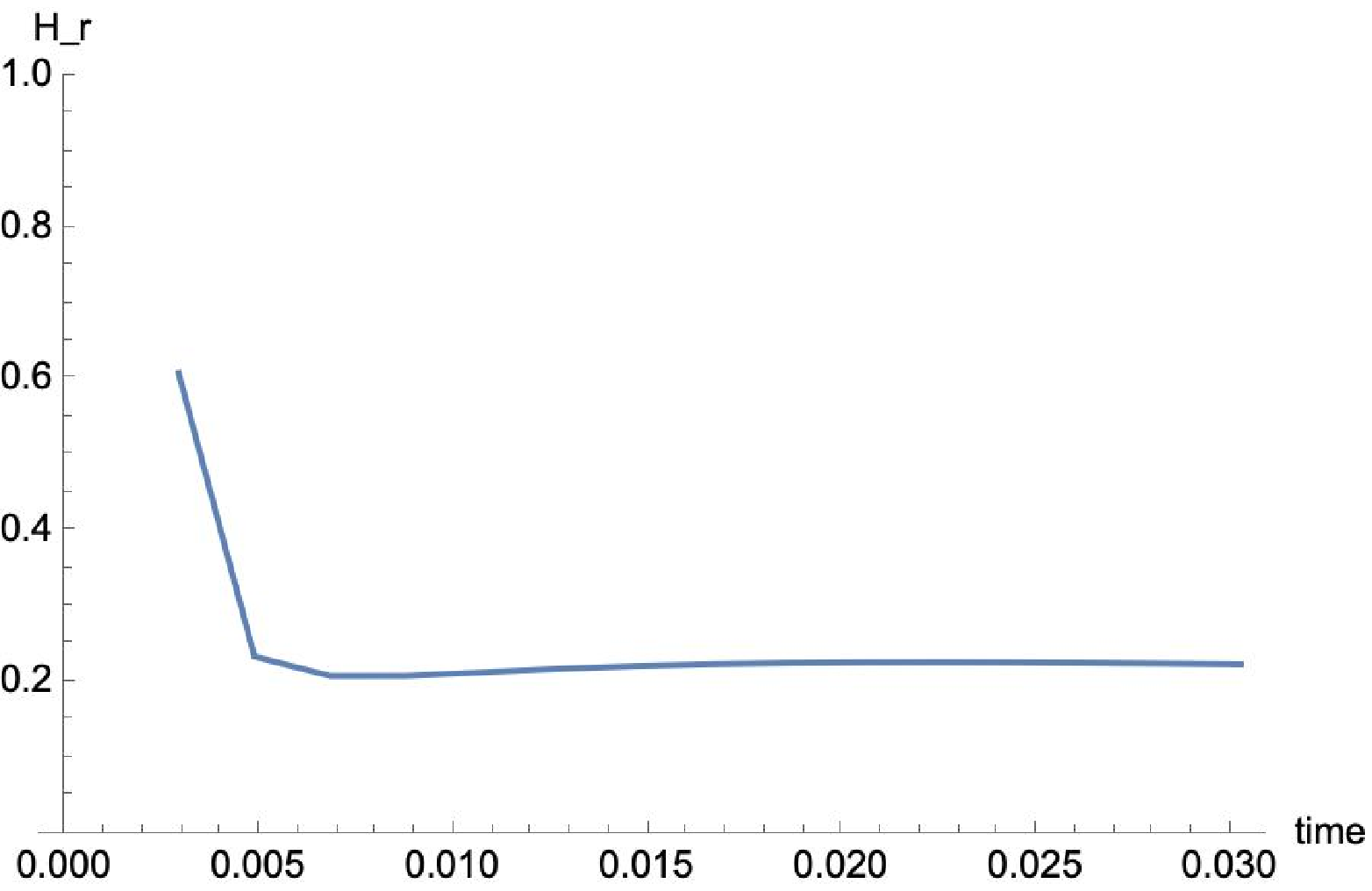}
			\subcaption{$H_r$}
			\label{3DHr}
		\end{minipage} \hfill&
		\begin{minipage}[t]{0.45\hsize}
			\centering
			\includegraphics[keepaspectratio, scale=0.4]{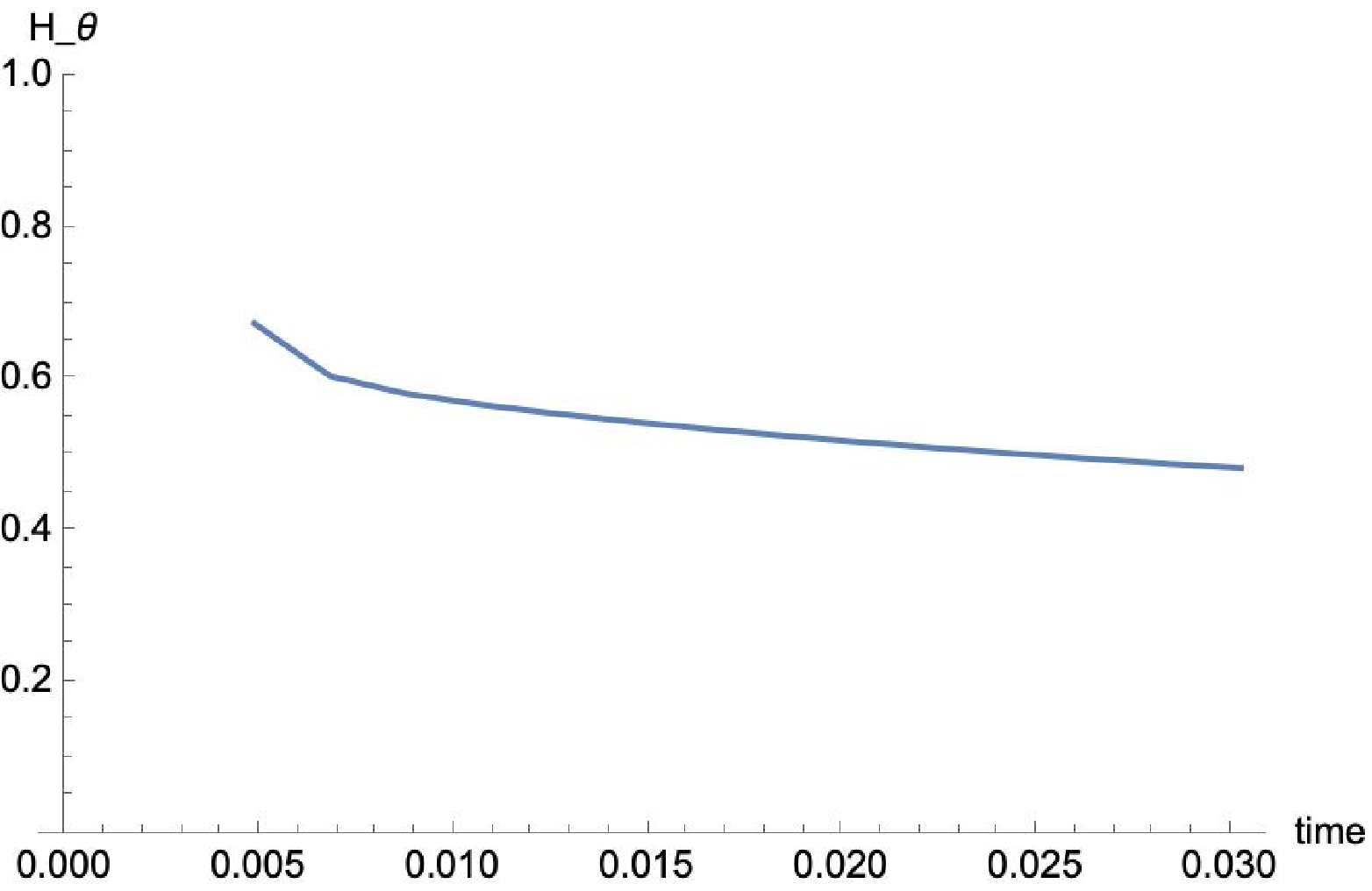}
			\subcaption{$H_\theta$}
			\label{3DHt}
		\end{minipage}
	\end{tabular}
	\caption{$H_r$ and $H_\theta$ for $D=3$. Horizontal axis describes time $t$.}
	\label{fig:D3H}
\end{figure}

We evaluate $H$, $H_r$, and $H_\theta$ numerically by taking the difference, for instance $H=\frac{1}{2}\frac{d}{dt}\ln\langle (\Delta \vec r)^2\rangle$,
which may depend on time. In Figure \ref{fig:D2H} (resp. Figure \ref{fig:D3H}), we illustrate the values of $H_r$, $H_\theta$ for $D=2$ (resp. $D=3$)\footnote{We have restricted the analysis in the range $0\leq t<0.0.3$. For the larger $t$, the distribution probability is not localized, and the Hurst exponents become ill-defined.}. 
For $D=2$, Hurst exponents vary in the range $0.6<H_r<0.8$ and $0.4< H_\theta<0.6$. For $D=3$, if we ignore the anomalous value in the beginning, the variation range is $H_r\sim 0.22$ and $0.5<H_\theta<0.6$.
As expected, Hurst exponents are different in the radial and angular directions, especially for $D=3$.

\begin{figure}[btp]
 	\begin{tabular}{cc}
 		\begin{minipage}[t]{0.45\hsize}
 			\centering
 			\includegraphics[keepaspectratio, scale=0.4]{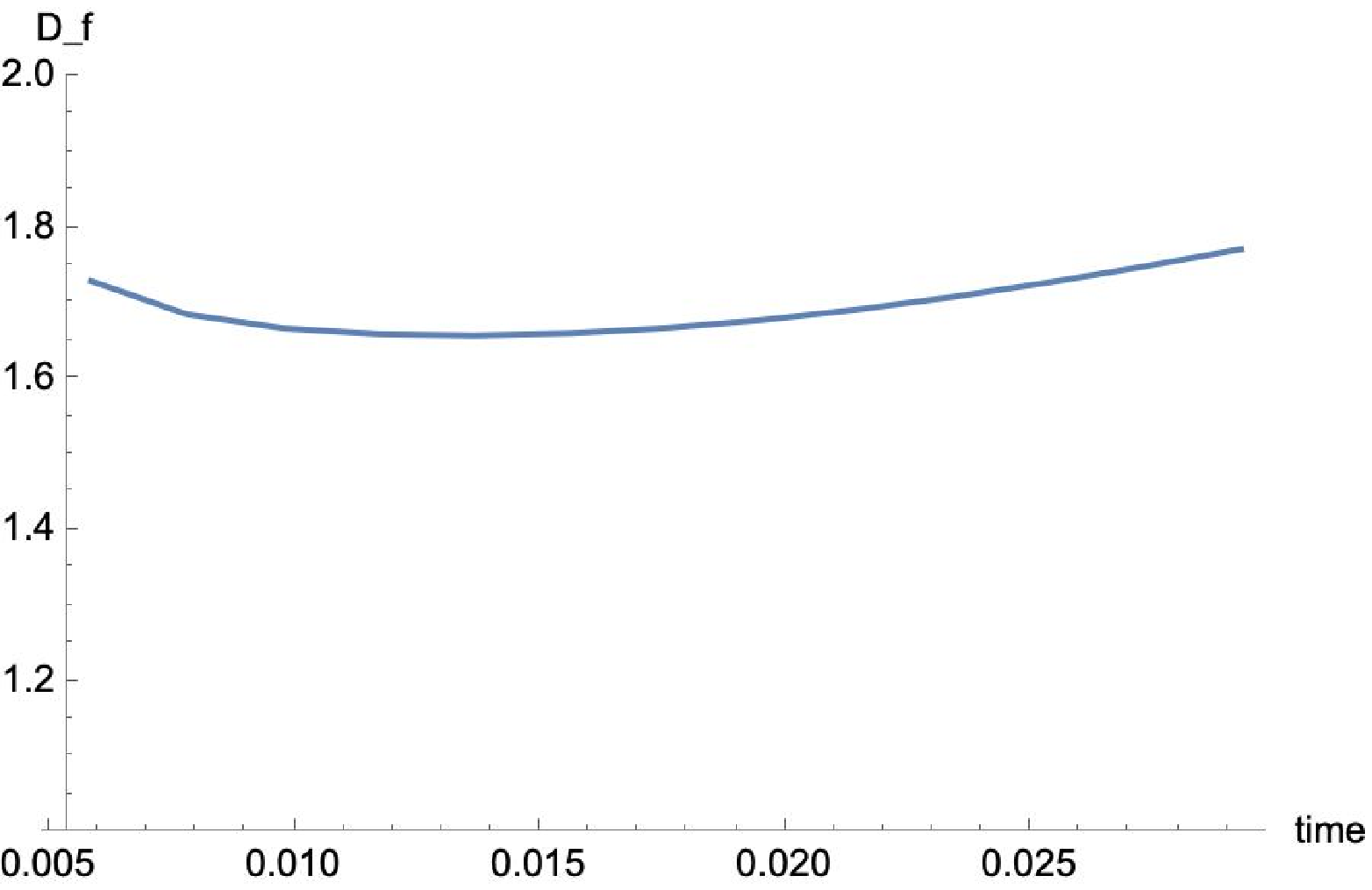}
 			\subcaption{$D_f$ for $D=2$}
 			\label{2DDf}
 		\end{minipage} \hfill&
 		\begin{minipage}[t]{0.45\hsize}
 			\centering
 			\includegraphics[keepaspectratio, scale=0.4]{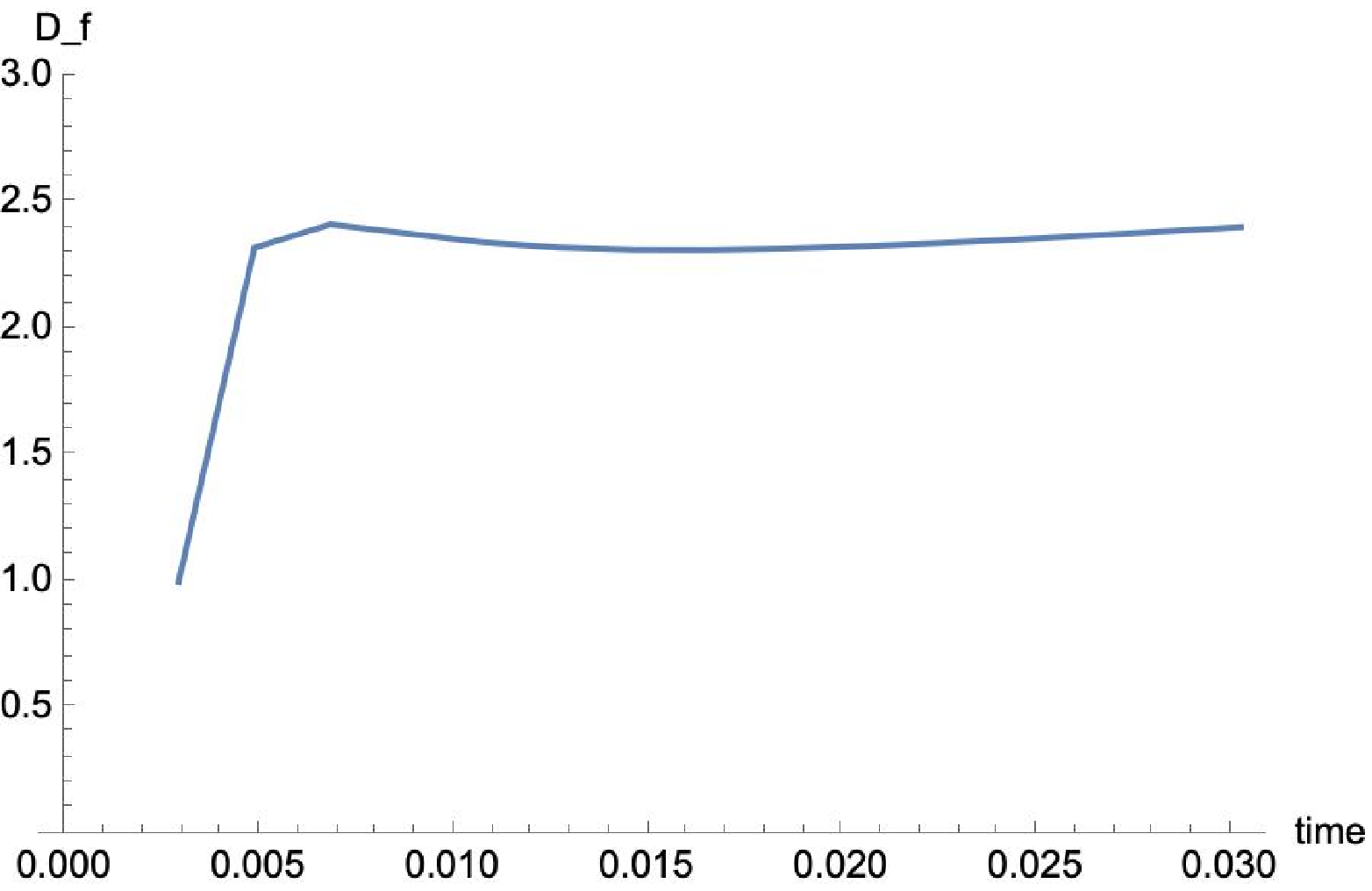}
 			\subcaption{$D_f$ for $D=3$}
 			\label{3DDf}
 		\end{minipage}
 	\end{tabular}
 	\caption{Fractal dimensions for $D=2, 3$. Horizontal axis describes time $t$.}
 	\label{fig:Df}
 \end{figure}

 Finally we give the plot of the fractal dimension $1/H$ in Figure \ref{fig:Df}. The obtained values, $D_f\sim 1.7$ for $D=2$ and $D_f\sim 2.4$, match with \cite{AFKMMSSYY}, at least in the range we studied\footnote{In the case of setups with directional dependence and multifractal behaviour in both radial and angular directions, there can be ambiguity in the definition of the fractal dimension, as summarised in Appendix D.}.
 
\subsection{Discussion of cut-offs with $r=\epsilon$}
\begin{figure}[btp]
    \begin{tabular}{cc}
      \begin{minipage}[t]{0.45\hsize}
        \centering
        \includegraphics[scale=0.3]{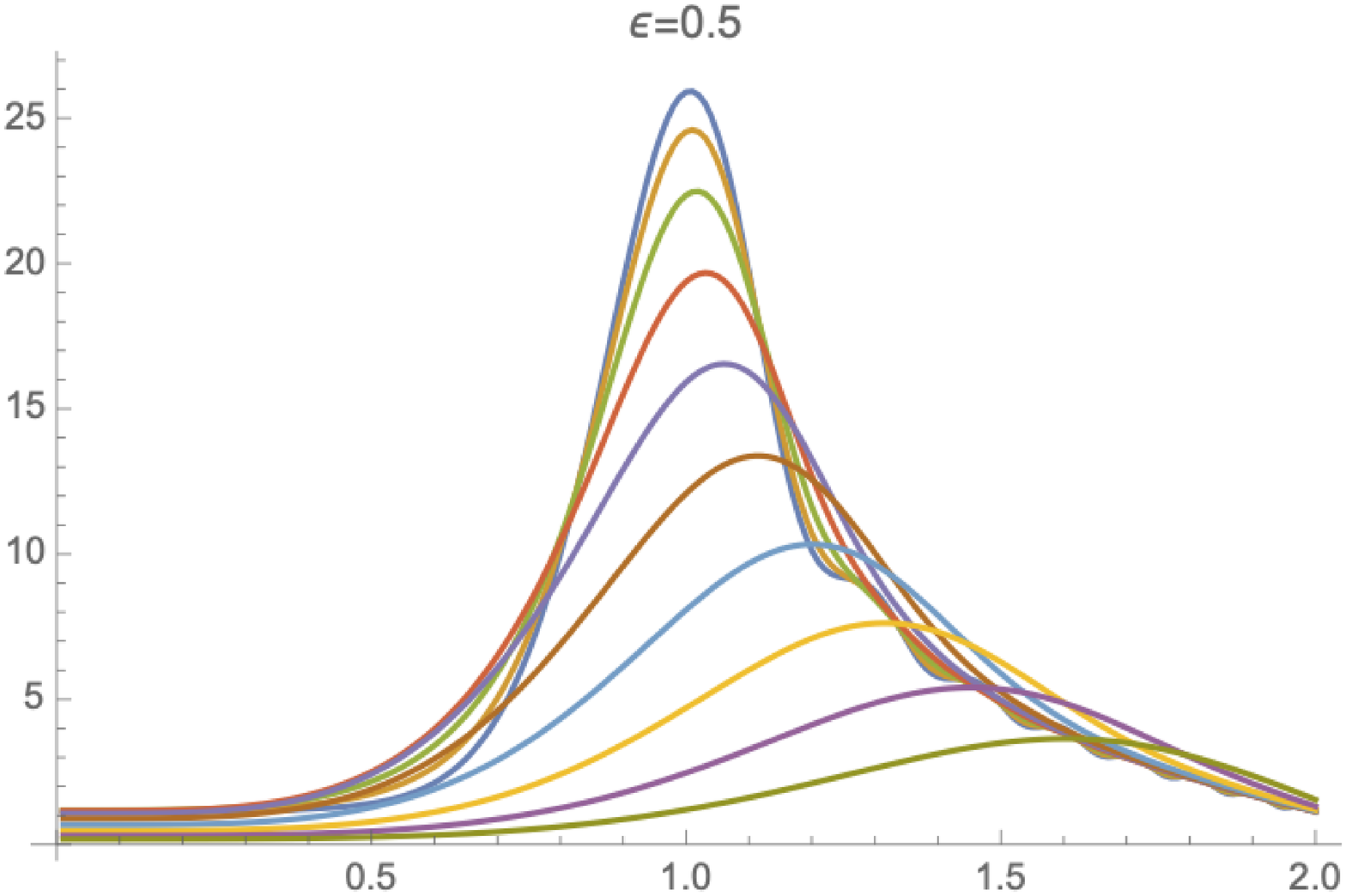}
        \subcaption{$D$=2}
        \label{radial_pD2_minus}
      \end{minipage} &
      \begin{minipage}[t]{0.45\hsize}
        \centering
        \includegraphics[scale=0.3]{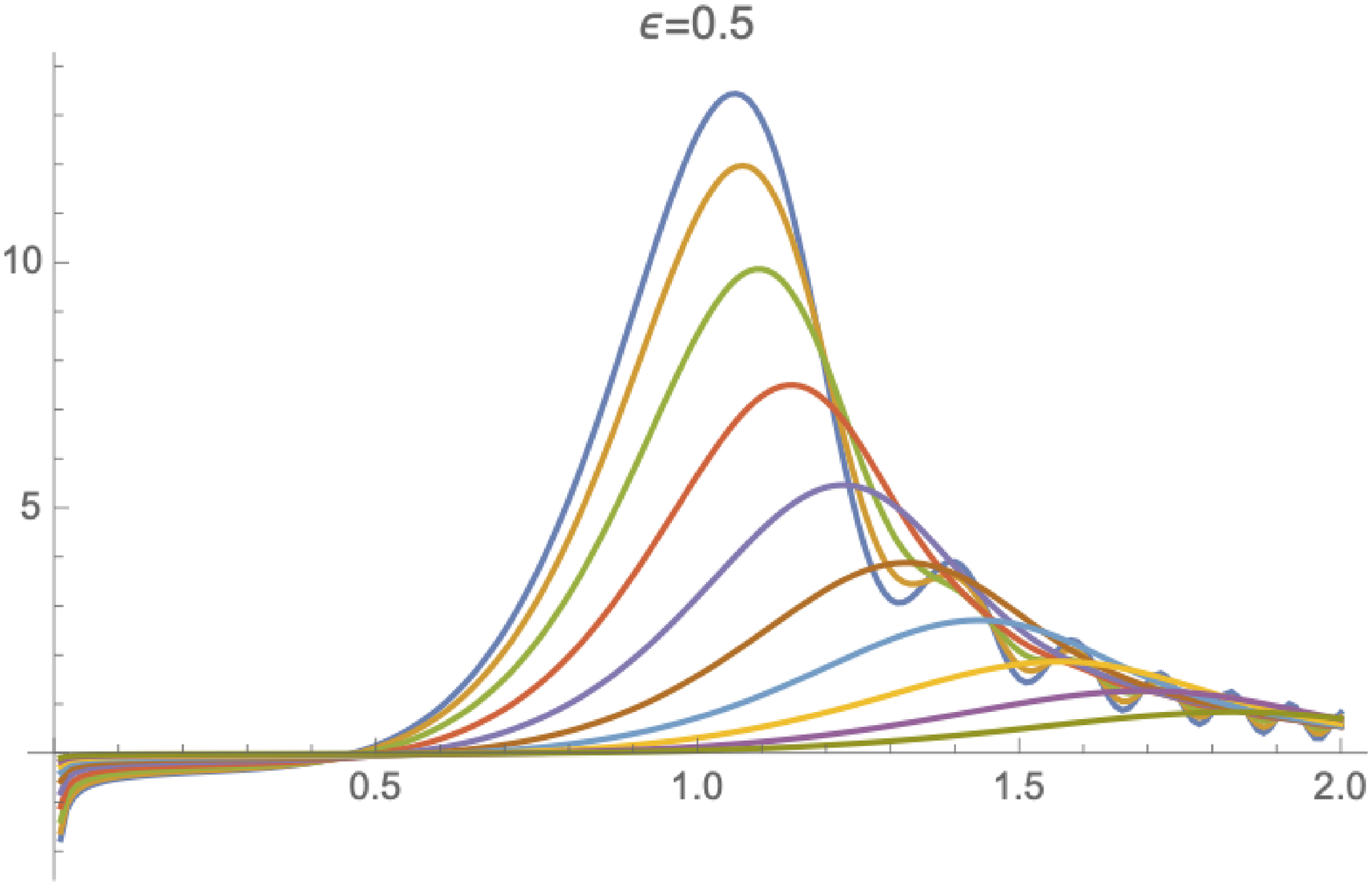}
        \subcaption{$D=3$}
        \label{radial_pD3_minus}
      \end{minipage} 
    \end{tabular}
     \caption{Shapes of $G(r,t)$ for various $t$. Horizontal axis gives the radius in case of cut-off $\epsilon=0.5$.}
     \label{fig:RadialProfileMinus}
  \end{figure}

\begin{figure}[btp]
 	\begin{tabular}{cc}
 		\begin{minipage}[t]{0.45\hsize}
 			\centering
 			\includegraphics[keepaspectratio, scale=0.4]{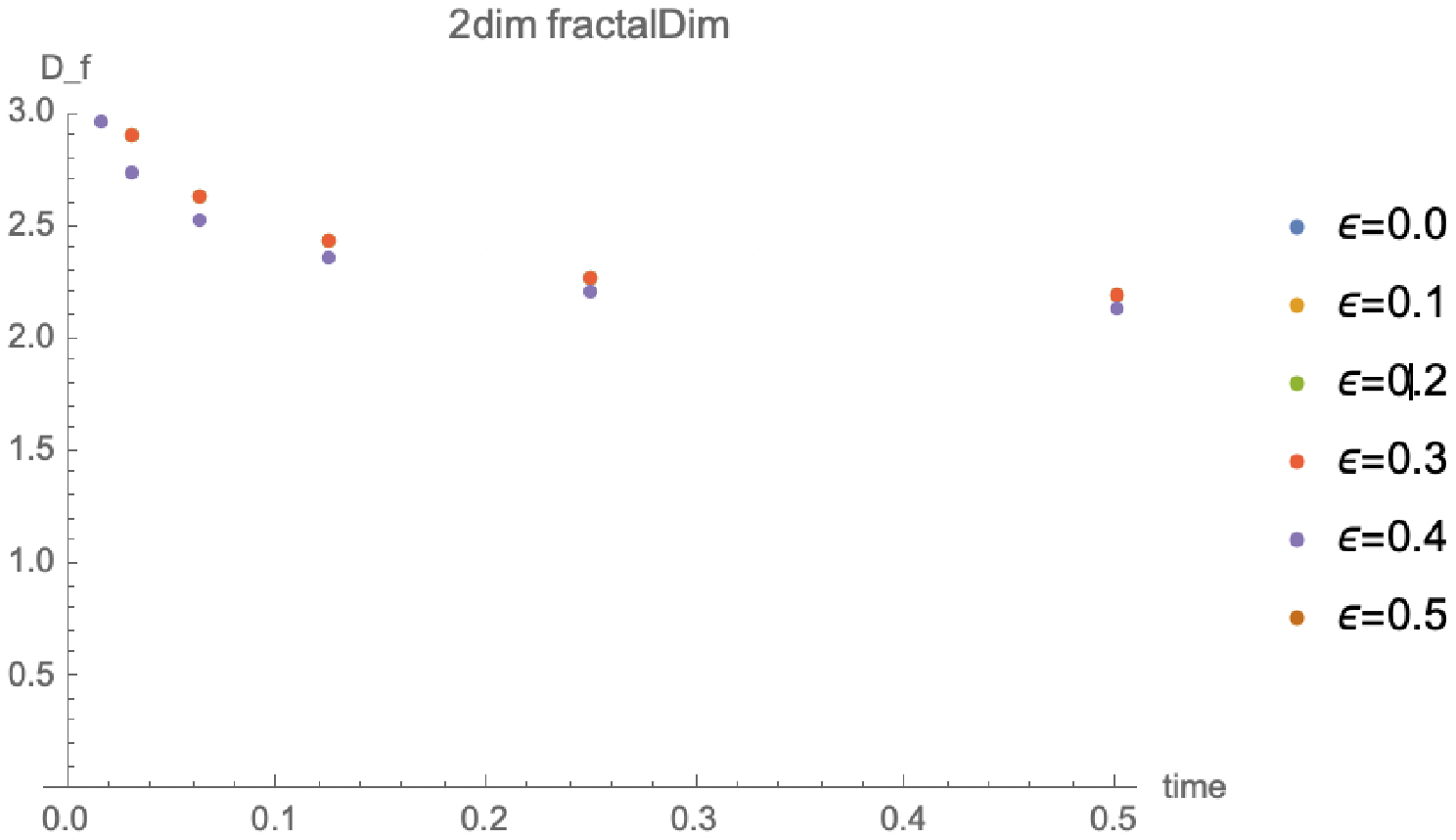}
 			\subcaption{$D_f$ for $D=2$}
 			\label{minus2DDf}
 		\end{minipage} \hfill&
 		\begin{minipage}[t]{0.45\hsize}
 			\centering
 			\includegraphics[keepaspectratio, scale=0.4]{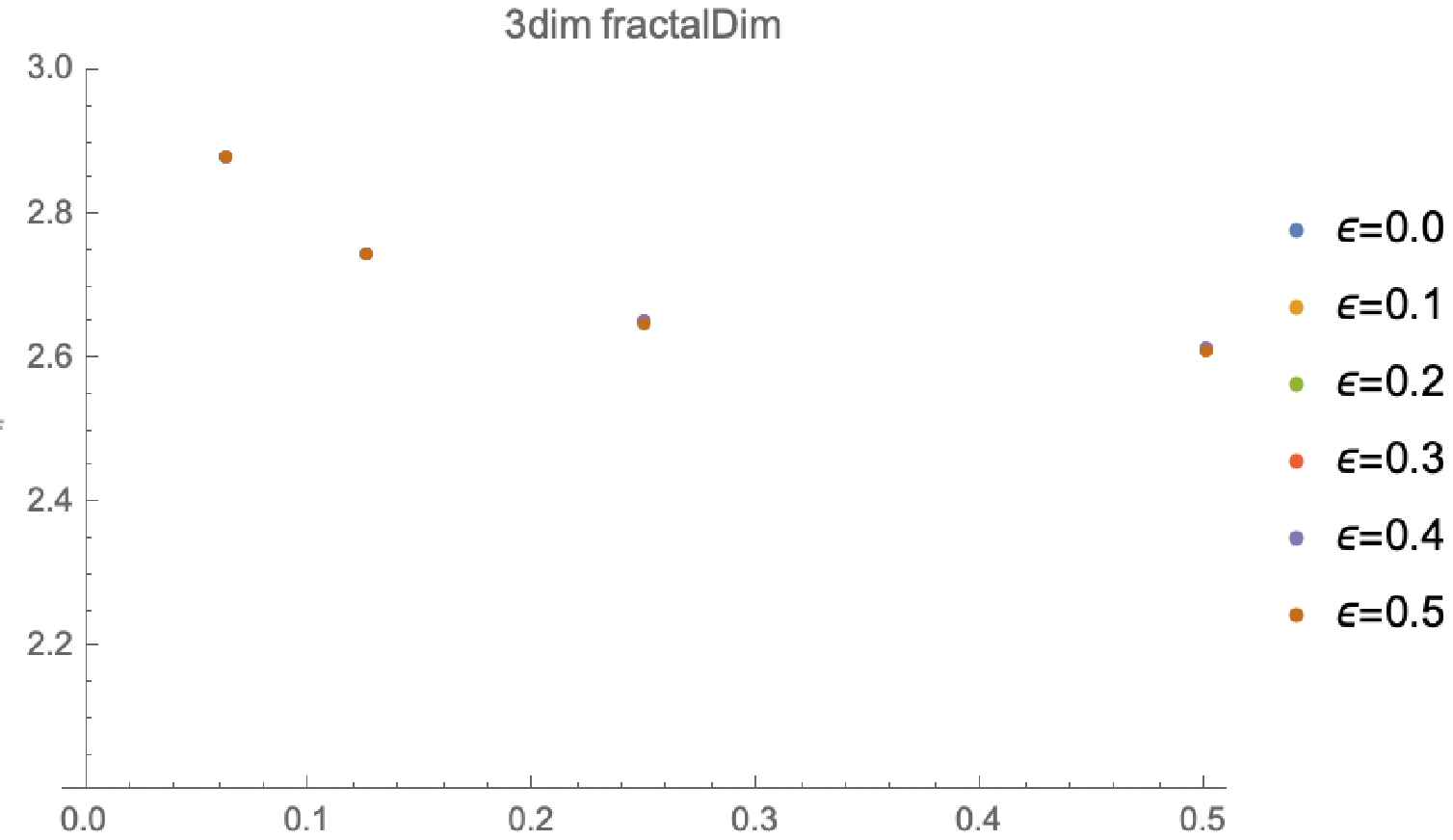}
 			\subcaption{$D_f$ for $D=3$}
 			\label{minus3DDf}
 		\end{minipage}
 	\end{tabular}
 	\caption{Fractal dimensions for $D=2, 3$ for varying cut-off $\epsilon$.}
 	\label{fig:DfMinus}
 \end{figure}

In this paper we discussed the singularity of $r=0$. Since this singularity can be artificial, and we try to set a boundary condition at $r=\epsilon$, which is close enough to the origin, such that there is no absorption of the particle, which corresponds to (\ref{BCD}). 
\begin{equation}
\left.r^{-D-1}\partial_{r}P(r,\vec{\omega},t)\right|_{r=\epsilon}=0.
\end{equation}
In this case, the term $J_{-\nu}$ also needs to be included in the solution, and the coefficients in (\ref{Rr}) are modified as 
\begin{eqnarray}
c_{-}&=&c_{+}A_{\mathrm{2dim}}(k,n)\epsilon^{\delta}, \quad\delta=\sqrt{n^2+4}\\
A_{2dim}(k,n)&=&\frac{6^{-\frac{2}{3}\delta}\left(\delta+2\right)k^{\frac{2\delta}{3}}\Gamma\left(1-\frac{\delta}{3}\right)}{\left(\delta-2\right)\Gamma\left(\frac{\delta}{3}+1\right)}
\end{eqnarray}
in the two-dimensions, and 
\begin{eqnarray}
c_{-}&=&c_{+}A_{3dim}(k,l)\epsilon^{\delta}, \quad \delta=\sqrt{l^2+l+25}\\
A_{3dim}(k,l)&=&-\frac{2^{-\delta}\left(\delta+5\right)k^{\frac{1}{4}\delta}\Gamma\left(-\frac{1}{8}\delta\right)}{\left(\delta-5\right)\Gamma\left(\frac{1}{8}\delta\right)}\,.
\end{eqnarray}
The constant $A_{2dim}(k,n)$ and $A_{3dim}(k,l)$ take into account of the $\epsilon$-dependence, sign, and the power of the cut-off of the mode dependence of the angular momentum. The solution and the numerically determined fractal dimension are plotted in Figure \ref{fig:RadialProfileMinus} and Figure \ref{fig:DfMinus} with $\epsilon=0.0\sim 0.5$.

The solution is influenced by the effect of the $J_\nu$ term near and far from the origin, and the fractal dimension takes large values for short times. This may indicate that the effect of $J_\nu$ term is a rapid spreading effect due to the bouncing off by the origin. In addition, the fractal dimension asymptotically approaches a range of values similar to the fractal dimension already shown. For instance, one may compare the behavior of the fractal dimension shown in Figure (\ref{fig:Df}) with that shown in Figure \ref{fig:DfMinus}.

\section{Conclusion}
In this paper, we derived the Fokker-Plank equation associated with the stochastic differential equation (\ref{eom}). We show that the system has an unusual property in that the total probability decreases with time, which is related to the jump of the particle in the vicinity of the dipole. We studied the properties of Green's function and found the modification necessary to compensate for the loss of probability, whose procedure depends on the prescription. We also noted the distinct scaling properties in the radial and angular directions. The fractal dimension, numerically observed in \cite{AFKMMSSYY} is reproduced by combining these ideas. We hope that the dipole system, which we studied, gives an interesting example of fractional diffusion, where the loss of the probability at the singularity gives rise to a fractal behavior of the particle's trajectory.

\section*{Acknowledgement}
We would like to thank Prof. Fukumoto for the invitation to the workshop, "Helicity and space-time symmetry -- a new perspective of classical and quantum systems ", October 5-8, 2021, OCAMI, Osaka City University, where a part of the result was announced.
YM is partially supported by JSPS Grant-in-Aid KAKENHI (\#18K03610, JP21H05190).

\appendix
\section{SLE, Bessel process and the Streamline by Randomly Modulated
Dipole}\label{SLEBessel}

In this section we discuss the relationship between our dipole model
and the SLE. The key to this discussion is the relation between the
SLE and the Bessel process.
A Bessel process is a stochastic process 
\begin{equation}
dr=\sqrt{\kappa}dB+\frac{D-1}{2}\kappa\frac{dt}{r}
\end{equation}
in the radial direction of a $D$-dimensional Brownian motion 
\begin{equation}
dr_{i}=\sqrt{\kappa}dB_{i},
\end{equation}
where $dB,dB_{i}$ is a Wiener process, satisfying the Ito rule 
\begin{equation}
dBdB=dt,\ dB_{i}dB_{j}=\delta_{ij}dt,
\end{equation}
and $\kappa$ is the diffusion coefficient.

By complexifying $r$ and performing a variable transformation as
in 

\begin{equation}
f_{t}(z)=z-\sqrt{\kappa}B,
\end{equation}
we obtain 
\begin{equation}
df_{t}(z)=\frac{\frac{D-1}{2}\kappa}{f_{t}(z)+\sqrt{\kappa}B}.
\end{equation}
If $\kappa$ is determined to take the numerator to 2 in order to
get the standard form,

\begin{equation}
\kappa=\frac{4}{D-1}
\end{equation}
the SLE equation 
\begin{equation}
df_{t}(z)=\frac{2}{f_{t}(z)+\sqrt{\kappa}B}.
\end{equation}
is obtained.

There is a relation 
\begin{equation}
\text{c = \ensuremath{\frac{(8-3\kappa)(\kappa-6)}{2\kappa}}}
\end{equation}
 between the $\kappa$ of the SLE and the central charge $\text{c}$ of the
conformal field theory (for details see \cite{SLECFT}). It is also related to the fractal dimension
$\delta$ by 
\begin{equation}
\delta=\begin{cases}
1+\frac{\kappa}{8} & 0\leq\kappa\leq8,\\
2 & \kappa\geq8.
\end{cases}
\end{equation}

Our dipole model is 
\begin{equation}
dr_{i}=d_{H}\sum_{j}V_{ij}^{(D)}dB_{j}
\end{equation}
 as a stochastic process. From the Ito rule, the second power of this
expression is
\begin{equation}
dr_{i}dr_{j}=d_{H}^{2}\sum_{k}V_{ik}^{(D)}V_{kj}^{(D)}dt.
\end{equation}

\begin{equation}
V_{ij}^{(D)}V_{jk}^{(D)}=\frac{1}{r^{2D}}(\delta_{ij}+D(D-2)\hat{r}_{i}\hat{r}_{j})
\end{equation}
Using these\footnote{
\begin{equation}
dr=\sum_{i}\frac{\partial r}{\partial r_{i}}dr_{i}+\frac{1}{2}\sum_{ij}\frac{\partial^{2}r}{\partial r_{i}\partial r_{j}}dr_{i}dr_{j}
\end{equation}

\begin{equation}
=d_{H}\sum_{i,j}\frac{r_{i}}{r}V_{ij}^{(D)}dB_{j}+\frac{d_{H}^{2}}{2}\sum_{ijk}\frac{1}{r}\left(-\hat{r}_{i}\hat{r}_{j}+\delta_{ij}\right)V_{ik}^{(D)}V_{kj}^{(D)}dt
\end{equation}
}, the radial stochastic process is

\begin{equation}
dr=-\left(D-1\right)\frac{d_{H}}{r^{D}}\sum_{j}\hat{r}_{j}dB_{j}+\frac{(D-1)}{2}\left(\frac{d_{H}}{r^{D}}\right)^{2}\frac{dt}{r}
\end{equation}

Using,

\begin{equation}
\kappa(r)=\left(\frac{d_{H}}{r^{D}}\right)^{2}
\end{equation}

the equation is
\begin{equation}
dr=-\sqrt{\kappa(r)}(D-1)dB+\frac{(D-1)}{2}\kappa(r)\frac{dt}{r}
\end{equation}

\begin{equation}
dB\equiv\sum_{i}\hat{r}_{i}\cdot dB_{i}
\end{equation}

This equation can be regarded as a Bessel process with the diffusion
coefficient depending on the position.
By complexifying $r$ and performing a variable transformation as
in 
\begin{equation}
f_{t}(z)=z+(D-1)\int\sqrt{\kappa(z)}dB
\end{equation}
we obtain 
\begin{equation}
df_{t}(z)=\frac{\frac{D-1}{2}\kappa(z)^{2}}{f_{t}(z)-(D-1)\int\sqrt{\kappa}dB}dt
\end{equation}
In summary, the dipole model is a generalization of the diffusion
coefficient in SLE with spatial dependence.

\section{Dirac quantization including variable $\lambda$}\label{DiracQ}
In the text, we used Polyakov's approach to derive the Fokker-Planck equation.
It may be of some interest to  treat the auxiliary field $\lambda$ as a dynamical variable by using Dirac bracket\cite{r:Dirac}.  We start from the action (\ref{S2}). In this section, we put $\dip=1$ for simplicity. The canonical variables are,
\begin{align}
	p_i &= \frac{\partial L_2}{\partial \dot r^i}=\lambda (V^{-2})_{ij}\frac{dr^j}{dt}\\
	p_\lambda &= \frac{\partial L_2}{\partial \dot \lambda}=0\,.
\end{align}
It implies that
\begin{equation}
	\phi_1\equiv p_\lambda \approx 0\,.\label{phi1}
\end{equation}
is the primary constraint. The symbol $\approx$ implies the equality in the weak sense. 
We first set the Poisson bracket for the dynamical variables by,
\begin{equation}
	\left\{r^i, p_j\right\}=\delta^i_j\,,\quad
	\left\{\lambda,p_\lambda\right\}=1\,.
\end{equation}
The Hamiltonian of the system is,
\begin{equation}
	H=p_i \frac{dr^i}{dt} -L_2=\frac{1}{2\lambda}(V^{-2})_{ij}p^i p^j+\frac{\lambda}{2}
\end{equation}
We have to check the consistency of the constraint (\ref{phi1}) by,
\begin{equation}
	\dot p_\lambda =\left\{p_\lambda, H\right\}=\frac12\left(
	\frac{1}{\lambda^2}(V^{-2})_{ij}p_ip_j-1
	\right)\approx 0\,.
\end{equation}
We find the secondary constraint,
\begin{equation}
	\phi_2=\frac{1}{\lambda^2}(V^{-2})_{ij}p_i p_j-1\approx 0\,.
\end{equation}

We note that the two constraints are not commutative,
\begin{equation}
	\left\{\phi_1, \phi_2\right\}=\frac{2}{\lambda^3} (V^{-2})_{ij}p_i p_j\approx \frac{2}{\lambda}\,.
\end{equation}
It implies that we have to modify Poisson bracket by Dirac bracket,
\begin{align}
	\left\{A, B\right\}_D&\equiv \left\{A, B\right\}+\left\{\phi_1,\phi_2\right\}^{-1}\left(\left\{A, \phi_1\right\}\left\{\phi_2, B\right\}-\left\{A, \phi_2\right\}\left\{\phi_1, B\right\}\right)\\
	&= \left\{A, B\right\}+\frac{\lambda}{2} \left(\left\{A, \phi_1\right\}\left\{\phi_2, B\right\}-\left\{A, \phi_2\right\}\left\{\phi_1, B\right\}\right)
\end{align}
After we replace Poisson bracket by Dirac bracket, it is clear that we have no more constraints.

We compute the Dirac bracket between the dynamical variables as,
\begin{align}
	& \left\{p_\lambda, \mbox{anything}\right\}_D=0\\
	&\left\{r^i, \lambda\right\}_D=\frac{1}{\lambda}(V^{-2})_{ij}p_j\\
	&\left\{p_i, \lambda\right\}_D=-\frac{1}{2\lambda}\partial_i (V^{-2})_{jk}p_j p_k\\
	&\left\{r^i,p_j\right\}_D=\delta^i_j\, ,\, \left\{r^i,r^j\right\}_D=\left\{p_i,p_j\right\}_D=0
\end{align}

The Hamiltonian, after the use of such constraint becomes,
\begin{equation}
	H=\lambda + u_1 \phi_1 + u_2 \phi_2 \,,
\end{equation}
where $u_1, u_2$ are arbitrary functions.
It looks too simple but one may confirm,
\begin{align}
	\frac{dr^i}{dt} &=\left\{ r^i, H\right\}_D=\frac{1}{\lambda}(V^{-2})^{ij} p_j,\\
	\frac{dp_i}{dt}& = \frac{1}{2\lambda} p_j \partial_i (V^{-2})^{jk} p_k\,,\\
	\frac{d\lambda}{dt}&=0\,,
\end{align}
which are equivalent to the classical equation of motions.
The last equation implies that one may put $\lambda$ to be a constant, which is consistent with Polyakov's analysis.

\section{Derivation of Green's function}\label{s:Green}
\subsection{Spherically symmetric case for general $D$}
We consider the solution that does not depend on the angular variables. 
For the initial value condition, we set $P(r,t=0)=p(r)$. 
We write the general solution in the integral form,
\begin{equation}\label{spherical}
P(r)=\int_0^\infty dk \rho(k) e^{-hk^2t/2}\left(\frac{r^{D+1}}{D^2-1}\right)^{\nu_D} J_{\nu_D}\left(\frac{kr^{D+1}}{D^2-1}\right)\,,
\end{equation}
where $\nu_D$ is given as (\ref{nuD}).
For the initial condition, it gives,
\begin{equation}
p(r)=\int_0^\infty dk \rho(k) \left(\frac{r^{D+1}}{D^2-1}\right)^{\nu_D} J_{\nu_D}\left(\frac{kr^{D+1}}{D^2-1}\right)\,.
\end{equation}
It is convenient to introduce a new variable,
$
\zeta=\frac{r^{D+1}}{D^2-1}
$, and we denote $q(\zeta)=p\left(r(\zeta)\right)$.
One may rewrite it in the form of the Hankel transformation (\ref{Hankel}),
\begin{align}
q(\zeta) \zeta^{-\nu_D}&= \int_0^\infty kdk \,\rho(k) k^{\nu_D-1} \,J_{\nu_D}(k\zeta)\label{eH1}\\
\rho(k) k^{\nu_D-1} &=\int \zeta d\zeta \,q(\zeta)\zeta^{-\nu_D} \,J_{\nu_D}(k\zeta)\label{eH2}
\end{align}
By rewriting the equation in the original variable $r$, the second equation becomes,
\begin{equation}
\rho(k) = \frac{1}{D-1} \left(\frac{k}{D^2-1}\right)^{\nu_D} \int_0^\infty  r^{3D/2} dr J_{\nu_D}\left(\frac{kr^{D+1}}{D^2-1}\right)p(r)\,.
\end{equation}
For $p(r)=\delta(r-1)$, 
\begin{equation}
\rho(k)=\frac{1}{D-1} \left(\frac{k}{D^2-1}\right)^{\nu_D} J_{\nu_D}\left(\frac{kr^{D+1}}{D^2-1}\right)\,.
\end{equation}
Plug this formula into (\ref{spherical}) gives (\ref{GreenSpherical}).

\subsection{Green's function for $D=2$ with angle}
The general solution to (\ref{FPD}) becomes,
\begin{equation}\label{solution2D}
P(r,\theta,t)=\int_0^\infty dk \sum_{n\in \mathbb{Z}} \rho_n(k) e^{
-\frac{htk^2}{2} } \left(\frac{kr^3}{3}\right)^{2/3} J_{\nu_n}\left(\frac{kr^3}{3}\right) e^{\mathrm{i}n\theta}.
\end{equation}
The coefficient $\rho_n(k)$ is fixed by the initial condition,
\begin{equation}
P(r,\theta,0)=P_0(r,\theta)=\frac{1}{2\pi}\int_0^\infty dk \sum_{n\in \mathbb{Z}} \rho_n(k)  \left(\frac{kr^3}{3}\right)^{2/3} J_{\nu_n}\left(\frac{kr^3}{3}\right) e^{\mathrm{i}n\theta}.
\end{equation}
To obtain $\rho_n(k)$, we first use the Fourier integral,
\begin{equation}\label{step1}
\int_{0}^{2\pi} d\theta \,e^{-\mathrm{i}n\theta} P_0(r,\theta) =p_n(r)=\int_0^\infty dk  \rho_n(k)  \left(\frac{kr^3}{3}\right)^{2/3} J_{\nu_n}\left(\frac{kr^3}{3}\right) \,.
\end{equation}

$\rho_n(k)$ is determined from the Hankel transformation as in the previous subsection with $\zeta=r^3/3$.
\begin{align}
	\rho_n(k)
	&=(k/3)^{1/3}\int_0^\infty dr\, r^{3} \int_0^{2\pi} d\theta e^{-\mathrm{i}n\theta} P_0(r,\theta)J_{\nu_n}(kr^3/3)\label{rhonk}
\end{align}
Plug it into (\ref{solution2D}) gives the solution of the FP equation.

In particular, it will be useful to apply the formula to the initial condition,
\begin{equation}
P_0(r,t)=\delta(r-1)\delta(\theta)\,.
\end{equation}
The corresponding solution play the role of Green's function.
Eq.(\ref{rhonk}) gives,
\begin{equation}
\rho_n(k)=(k/3)^{1/3} J_{\nu_n}(k/3)\,.
\end{equation}
Eq.(\ref{solution2D}) gives (\ref{Green2D}).


 \subsection{Green's function for $D= 3$ with angle}
From the analysis of section \ref{s:DdimSOl}, the eigenfunction of $\mathcal{K}$ for $D=3$ is,
\begin{align}
&\mathcal{K}=r^{-8}\left(
 4 (r^2\partial_r^2-4r\partial_r)+\hat\Omega \right),\qquad
 \mathcal{K}\psi_{klm}=-k^2\psi_{klm}\\
&\psi_{klm}(r,\theta,\varphi,t) = r^{\frac52} J_{\nu_\ell} \left(\frac{kr^4}{8}\right)Y_{\ell m}(\theta,\varphi),\qquad
\nu_\ell=\frac{\sqrt{\ell(\ell+1)+25}}{8}\,,
\end{align}
where $Y_{lm}(\theta,\varphi)$ ($l=0,1,2,\cdots,$, $m=-l,-l+1,\cdots, l$) is the spherical harmonics satisfying $\hat\Omega Y_{lm}=-l(l+1) Y_{lm}$.

%
%
%
The general solution to (\ref{FPD}) is a linear combination of them,
\begin{equation}\label{solution3D}
	P(r,\theta,\varphi,t)=\int_0^\infty dk\sum_{l,m} \rho_{lm}(k)e^{-\frac{htk^2}{2}}\left(\frac{kr^4}{8}\right)^{5/8} J_{\nu_l}(\frac{kr^4}{8})Y_{lm}(\theta,\varphi) 
\end{equation}
where $\rho_{lm}(k)$ is an arbitrary function, which is determined by the initial value problem.
\begin{equation}\label{init}
	P(r,\theta,\varphi,0)=P_0(r,\theta,\varphi)=\int_0^\infty dk\sum_{lm}\rho_{lm}(k)\left(\frac{kr^4}{4}\right)^{5/8} J_{\nu_l}(\frac{kr^4}{4})Y_{lm}(\theta,\varphi)\,.
\end{equation}
%
%
The derivation of $\rho_{lm}(k)$ from the initial data is completely parallel to the 2D case, ($d^2\omega=\sin\theta d\theta d\varphi$)
\begin{align}
	\rho_{lm}(k)&=(k/8)^{3/8} \int_0^\infty dr\, r^{9/2} \int d^2\omega P_0(r,\theta,\varphi)J_{\nu_l}(kr^4/8)Y^*_{lm}(\theta,\varphi)\,.\label{rholm}
\end{align}
To derive Green's function, one uses the initial condition localized at $r=1$ and $\theta=0$, namely the north pole.\footnote{To be precise, one may use a distribution $P_0(r,\theta,\varphi)=\frac{1}{2\pi}\delta(r-1)\hat\delta(\theta)$ with $\hat\delta(\theta)=0$ for $\theta\neq 0$ and $\int_0^\pi \sin\theta\hat\delta(\theta)=1$.} If we start from such an initial value, the solution should be homogeneous in $\varphi$ direction, which implies that $\rho_{lm}=0$ for $m\neq 0$. We use $Y_{l0}(0,\varphi)=\sqrt{\frac{2l+1}{4\pi}}$ to obtain the explicit formula for Green's function (\ref{green3D}).

\section{Fractal dimension}
In this appendix, we  discuss the fractal dimension\footnote{See \cite{HALVLIN}\cite{KANTELHARDT} for more details.}.

\subsection{Box Dimension}

Let $V=L^{D}$ be the whole space, and divide the space into sections
of length $l$. Let 
\begin{equation}
M=\frac{L^{D}}{l^{D}}
\end{equation}
be the number of divisions.
Suppose there are $Z$ points in total\footnote{These symbols are used in order to be aware of the similarities with
statistical mechanics.}. The number of points in the $i$-th cell is 
\begin{equation}
Z_{l}(i).
\end{equation}
The probability that a point is in the $i$-th cell is
\begin{equation}
p_{l}(i)=\frac{Z_{l}(i)}{Z}.
\end{equation}
Let 
\begin{equation}
Z_{l}=\sum_{i}\theta(p_{l}(i))
\end{equation}
be the number of cells if there is at least one point in a cell.
We assume the following power relation 
\begin{equation}
Z_{\mu l}=\mu^{F_{0}}Z_{l}
\end{equation}
where $F_{0}$ is called Box dimension( fractal dimension).
If we set $\mu\to l,l\to1$,
\begin{equation}
Z_{l}\propto l^{F_{0}}.
\end{equation}

\subsection{Case of position-dependent scale transformation}

If we take the size $l$ to be small enough, the distribution will
have a power relation.
\begin{equation}
p_{\lambda l}(i)=\lambda^{S(i)}p_{l}(i)
\end{equation}
where $S(i)$ is called the singularity exponent or Lipschitz-H\"{o}lder
exponent\footnote{This S, together with T, which appears next, plays the role of entropy and temperature in thermodynamics.}.
As in the previous section, we introduce 
\begin{equation}
p_{l}(i)\propto l^{S(i)},
\end{equation}

\begin{equation}
Z_{l}(i)\propto l^{S(i)}.
\end{equation}
We now define 
\begin{equation}
B_{S}(l)=\{i|S(i)=S\}
\end{equation}
to be the region with the same scale dimension.
The number of elements of $B_{S}(l)$ is
\begin{equation}
Z_{S}(l)=\#B_{S}(l)=\sum_{\{i|p_{l}(i)\neq0\}}\delta(S(i)-S).
\end{equation}
From the above, $Z_{l}$ can be written as
\begin{equation}
Z_{l}=\int dSZ_{S}(l).
\end{equation}
If $Z_{S}(l)$ satisfies a power law, then using scale dimension $E(S)$,
\begin{equation}
Z_{S}(\lambda l)\equiv\lambda^{E(S)}Z_{S}(l).
\end{equation}
From this,
\begin{equation}
Z_{S}(l)\propto l^{E(S)},
\end{equation}

\begin{equation}
Z_{l}\propto\int dSl^{E(S)}.
\end{equation}
It follows that if $E(S)$ is constant, then it corresponds to the
box dimension $F_{0}$.
\begin{equation}
Z_{l}\propto l^{F_{0}}
\end{equation}

\subsection{MultiFractal dimension}

The box-counting we have seen so far ignores the possible fluctuations
(or intermittency) of the probability in a cell. In order to take
into account the stochastic contribution, we take into account the
``temperature''. The corresponding ``free energy'' is calculated.

For the $Z_{l}$ already defined, we introduce the following partition
function which coincides with $Z_{l}$ at $T=0$.

\begin{equation}
Z_{l}(T)\equiv\sum_{i}p_{l}(i)^{T}\theta(p_{l}(i))
\end{equation}

Let us assume that this partition function also has a power row, using
scale dimension $F(S)$, 

\begin{equation}
Z_{\lambda l}(T)=\lambda^{F(T)}Z_{l}(T).
\end{equation}

Then,
\begin{equation}
Z_{l}(T)=l^{F(T)}.
\end{equation}
From the definition, 
\begin{equation}
Z_{l}(T)=\sum_{i}a(i)l^{TS(i)}\theta(p_{l}(i))
\end{equation}

\begin{equation}
=\int dSl^{TS}\sum_{i}a(i)\delta(S(i)-S)\theta(p_{l}(i))=\int dSl^{TS}Z_{S}(l)
\end{equation}

\begin{equation}
=\int dSl^{TS-E(S)}
\end{equation}

where $a(i)$ is the weight as a coefficient of the power law.

Thus, by the saddle point approximation, the free energy is
\begin{equation}
F(T)=\min_{S}\left(TS-E(S)\right).
\end{equation}

In multifractal terms, $T$ is $q$ and the free energy is called
the generalised dimension;

\begin{equation}
F(q)=\frac{D_{q}}{q-1}
\end{equation}

\subsection{Correlation function and fractal dimension}

Here we explain how we calculate the fractal dimension used in this
paper.

Instead of the partition function, we have considered the two point
correlation function in this paper,
\begin{equation}
Z_{l}\to\langle\left(\Delta\vec{r}(t)\right)^{2}\rangle.
\end{equation}
Corresponding to the cells are the directional components; $i=r,\theta.$

The scale dimension is assumed to be different in different directions.
\begin{equation}
\Delta_{r}\propto t^{S_{r}}
\end{equation}

\begin{equation}
\Delta_{\theta}\propto t^{S_{\theta}}
\end{equation}

The temperature is introduced according to the multifractal technique,

\begin{align}
\langle\Delta\vec{r}(t)^{T}\rangle &=Z_{t}(T)=\sum_{i=r,\theta}a(i)t^{TS(i)}\theta(p_{i}(t)) \\
 &=\int dSt^{TS}\sum_{i=r,\theta}a(i)\delta(S(i)-S)\theta(p_{i}(t)) \\
&  =\int dSt^{TS}Z_{S}(t)=\int dSt^{TS-E(S)}.
\end{align}

From saddle point approximation, we obtain
\begin{equation}
F(T)=\min_{S}(TS-E(S)).
\end{equation}

In this paper, we consider the fractal dimension as a response of
spatial scales by the time scale transformations as

\begin{equation}
t=\Delta x_{T}{}^{\frac{1}{F(T)}}\equiv\Delta x_{T}^{D_{F}}
\end{equation}

\begin{equation}
\Delta x_{T}\equiv\langle\Delta\vec{r}(t)^{T}\rangle^{\frac{1}{T}}
\end{equation}
and $T=2$. In the paper, $S(i)$ is reffered to as $H_{i}$.

Note that the fractal dimension is defined differently from the fractal
dimension calculated in the companion paper. There, cells are space
separators and the mesh size $l$ is used instead of $t$.

Whether the dimension obtained from the correlation function coincides
with that obtained from the box counting method depends on the problem
to which we apply them, but both results coincide for normal fractal
structures such as Brownian motion.

\end{document}